\documentclass[aps,prb,floatfix,amsmath,showpacs,lengthcheck]{revtex4-1}
\usepackage{graphicx}
\usepackage{amssymb}
\usepackage{url}
\usepackage[colorlinks=true,%
urlcolor=blue,% 
linkcolor=blue,% 
citecolor=blue,% 
hypertexnames=true,%
]{hyperref}% 
\begin{document}
	
\title{Particle-In-Cell Simulation of Ultrafast Hot-Carrier Transport in Fe/Au-heterostructures}
	
\author{Dennis M. Nenno}
\affiliation{Physics Department and Research Center OPTIMAS, University of Kaiserslautern, 67663 Kaiserslautern, Germany}
\author{B\"{a}rbel Rethfeld}
\affiliation{Physics Department and Research Center OPTIMAS, University of Kaiserslautern, 67663 Kaiserslautern, Germany}
\author{Hans Christian Schneider}
\email{hcsch@physik.uni-kl.de}
\affiliation{Physics Department and Research Center OPTIMAS, University of Kaiserslautern, 67663 Kaiserslautern, Germany}


\date{\today}

\begin{abstract}
	We describe a theoretical approach for spin-polarized hot-electron transport, as it occurs after excitation by ultrafast optical pulses in heterostructures formed by ferromagnetic and normal metals. We formulate a spin-dependent particle-in-cell model that solves the Boltzmann equation for excited electrons. It includes lifetimes and transmission coefficients as parameters, which can be taken from \textit{ab initio} calculations or experiment, and can be easily extended to multilayer systems. This approach is capable of describing electron transport in the ballistic, super-diffusive and diffusive regime including secondary-carrier generation. We apply the model to optically excited carriers in Fe/Au bilayers and Fe/Au/Fe spin-valve structures. We gain microscopic insight into the hot-electron transport dynamics probed in recent experiments on spin-valves. We find contributions to the demagnetization dynamics induced in Fe/Au/Fe trilayers regardless of the parallel or antiparallel magnetic alignment of the Fe layers.
\end{abstract}

\maketitle

%%%%%%%%%%%%%%%%
% Introduction %
%%%%%%%%%%%%%%%%
\section{Introduction}
The excitation of ferromagnetic metals by ultrashort pulses and the subsequent ultrafast magnetization dynamics have been under investigation for two decades. In recent years, interest has shifted from the dynamics in homogeneous magnetic films to magnetic multilayers and magnetic nanostructures.  Experimental and theoretical results have inched the field closer to a microscopic understanding of the different processes involved in ultrafast magnetization dynamics, however it remains a complicated subject due to the variety of effects involved. An important process contributing to the magnetization/spin dynamics in metallic systems after ultrashort-pulse excitation is the transport of hot carriers, i.e., carriers excited far away from the Fermi energy. Initially, transport of these hot carriers is mainly ballistic and becomes diffusive due to carrier scattering events. The intermediate regime is sometimes called superdiffusive transport and its effect on the demagnetization process in optically excited ferromagnets was first analyzed theoretically by Battiato \textit{et al}.~\cite{Battiato:2010br} Their insight was that different velocities and lifetimes of excited carriers in both spin channels may lead to an effective change of the spin polarization of excited electrons and thus also on the magnetization. Signatures of this effect were subsequently seen in the magnetization dynamics of thin films.~\cite{Rudolf:1ky} Recent studies have supported the finding that it is indeed the hot-carrier dynamics that has an influence on the demagnetization in metallic multilayers~\cite{Bergeard:2016jka,Salvatella:2016gz,Turgut:2016dx} without a direct interaction with the laser field. This also excludes the possibility of demagnetization by coherent light-matter interactions.~\cite{bigot2009coherent}
	
Apart from its connection with the change of magnetization in ferromagnetic layers, hot-carrier dynamics also play an important role in the spin-dependent electronic transport in normal metals that can be measured, for instance, using nonlinear-optical techniques~\cite{Melnikov:2011ep}, or the C-MOKE technique~\cite{Schellekens2014,Hofherr:2017dya}. These novel optical techniques allow one to measure directly the depth-dependence of the carrier transport on ultrashort timescales. In order to understand the detailed spatio-temporal characteristics of carrier transport, further development of theoretical models is needed.
	
Standard electronic transport models, such as the linearized Boltzmann transport equation~\cite{Valet1993,Zhang2002,Xiao2007} and Kubo formulas~\cite{Butler1985,Zhang1992} are usually restricted to transport close to the Fermi surface where the currents are driven by longitudinal electric fields with typical modulation frequencies of a few GHz. In that case, it is a good approximation to assume that scattering processes lead to diffusive transport and the influence of the electric field changes the electronic distributions only around the Fermi energy. 
	
For the dynamics of hot carriers in metallic heterostructures, different theoretical approaches exist, ranging from macroscopic, extended wave-diffusion equations for transport close to the Fermi edge~\cite{Kaltenborn:2014hg} to many-particle Monte-Carlo models~\cite{Huthmacher:2016dd} for excitation at high energies up to 100\,eV. Recent experimental studies also make use of transport extensions of the three-temperature model.~\cite{Wieczorek:2015fk}
	
To model the effect of optically excited hot electrons, Battiato \textit{et al}.\ introduced classical equations of motion that describe the spin-dependent carrier population~\cite{Battiato:2010br} together with \textit{ab initio} input for carrier velocities and lifetimes. They showed that this model is capable of describing superdiffusive transport, i.e., the transition between ballistic and diffusive transport and applied to a number of recent experiments.~\cite{Kampfrath:2013kw,Eschenlohr:2013id} Due to its derivation from a phenomenological picture of carrier transport, the limits of its applicability, e.g., concerning the system size, cannot be simply understood. For the same reason, the model cannot be easily extended to more complex systems.~\cite{Battiato:2014kz}

In this paper, we present an approach based on the Boltzmann transport equation (BTE) for the hot-electron distribution function. This equation is numerically solved using the particle-in-cell (PIC) approach combined with an operator-splitting technique. We show that this method reproduces the transport characteristics from ballistic to diffusive on typical experimental timescales and includes the secondary-carrier cascade. We present results on Fe-Au bilayers and show how the transmission profile of the interface contributed to the polarization of injected carriers and spin-currents. In a Fe-Au-Fe spin-valve structure with parallel and antiparallel alignment, we find enhancement of the polarization above the equilibrium polarization. For antiparallel alignment of emitter and collector, however, we can show that this behavior is limited to a thin penetration depth inside the collector region. 
	
The paper is organized as follows. Our approach to set up a BTE for hot-electron transport is introduced in Sec.~\ref{sec:model}, followed by the numerical solution procedure outlined in Sec.~\ref{sec:numerics}. Some capabilities of the model are presented in Sec.~\ref{sec:model_studies}, before we turn to discuss our results on Fe-Au bilayers and Fe-Au-Fe structures in Sec.~\ref{sec:results}.
	
	%%%%%%%%%%
	% Theory %
	%%%%%%%%%%
\section{Model\label{sec:model}}
For the description of the spatio-temporal dynamics of laser-excited hot-electrons electrons, we start from the Boltzmann Transport Equation. As has been shown often (see, e.g., Ref.~\onlinecite{Silin:1957ko,Bonitz:2015up}), the BTE can be derived as the evolution equation of a reduced single-particle density matrix. Its approximations to the BTE have been used extensively to describe carrier dynamics in magnetic multilayers ~\cite{Qi:2003ks,Valet:1993es,Zhu:2009dm}. It has also proved to be a useful tool for the study of conduction-electron dynamics in thin metal layers ($\approx 10$\,nm).~\cite{Manfredi:2005ba,PhysRevB.97.014424}
	
We follow an approach that originated in the 1950s with the work of Wolff~\cite{Wolff:1954ds} to understand the electron cascade due to the interaction/scattering with neutrons or high-energy electrons. Penn \textit{et al}.~\cite{Penn:1985us} later used this method including the spin dependence to describe the polarization of non-equilibrium carriers in ferromagnets. As starting point we take the evolution equation of the carrier-distribution function in the form~\cite{Penn:1985wt}
	\begin{equation}
	\begin{split}
	\left[\partial_t + \frac{\hbar}{m}\mathbf{k}\cdot\mathbf{\nabla}_{\mathbf{r}}+\frac{1}{\hbar}\mathbf{F}_{\sigma}(\mathbf{r},t)\cdot\mathbf{\nabla}_{\mathbf{k}}\right] f_{\sigma}(\mathbf{r},\mathbf{k},t) = S_{\sigma}(\mathbf{r},\mathbf{k},t) \\- \frac{f_{\sigma}}{\tau^{\mathrm{eff}}_{\sigma}(\mathbf{r},E)} + \sum_{\sigma'}\int d^3k'w(\mathbf{r},\sigma',\mathbf{k'};\sigma,\mathbf{k})f_{\sigma'}(\mathbf{r},\mathbf{k}',t) \, ,
	\label{eq:bte}
	\end{split}
	\end{equation}
	where $f$ depends on position $\mathbf{r}$, crystal momentum~$\mathbf{k}$, time~$t$ and $S$ is a source term used to describe the excitation process. The last two terms in Eq.~\eqref{eq:bte} describe out- and in-scattering processes due to interactions with equilibrium carriers and many-particle excitations. In the form~\eqref{eq:bte}, the relaxation-time approximation was applied to the out-scattering term, and the effective, spin-dependent out-scattering time is given by
	\begin{equation} \tau^{\mathrm{eff}}_{\sigma}(\mathbf{r},E)=1/(\tau^{-1}_{\mathrm{el}}+\tau^{-1}_{\sigma}(\mathbf{r},E)) \, ,\label{eq:lifetimes}
	\end{equation}
	which combines the elastic lifetime $\tau_{\mathrm{el}}$ and the inelastic lifetime $\tau_{\sigma}(E)$. The relaxation times depend on the electron position in the material as well. The in-scattering amplitude $w$ describes transitions of electrons scattering from a state with spin $\sigma'$ and momentum $\mathbf{k}'$ into a state $\sigma$, $\mathbf{k}$.~\cite{Penn:1985wt} All scattering events are assumed to be local. For completeness, we have also included in Eq.~\eqref{eq:bte} a force contribution $\mathbf{F}_{\sigma}$ acting on electrons, which may be due to internal and/or external fields. However, we do not numerically evaluate this term in the present paper.
	
	% Figure 1
	\begin{figure}[t]
		\includegraphics[width=0.45\textwidth]{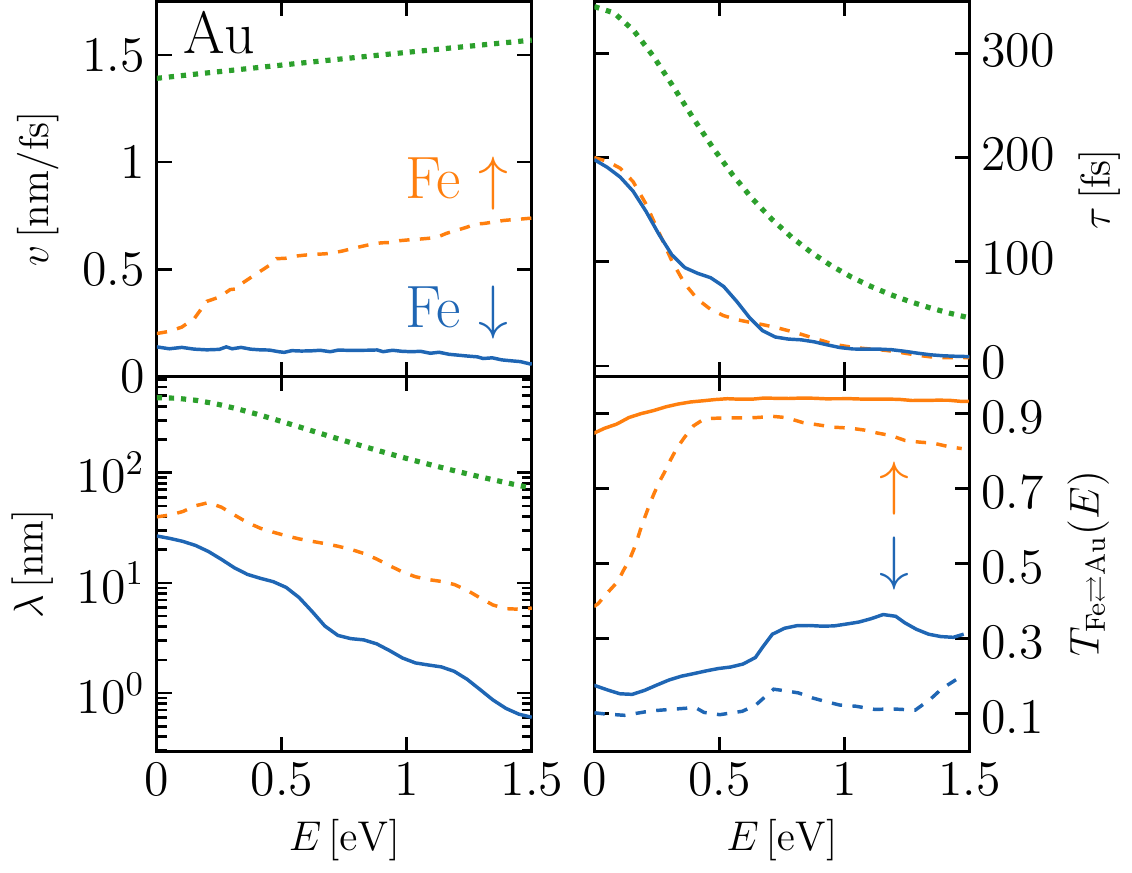}
		\caption{\textit{Ab initio} data used to simulate hot-carrier dynamics in bilayers and trilayers. Velocities, taken from Ref.~\onlinecite{Zhukov:2006ky}, inelastic lifetimes~\cite{Kaltenborn:2014du} and inelastic mean-free paths are shown for majority (dashed line) and minority (solid line) electrons in Fe and in Au (dotted line). The interface transmittance, taken from Ref.~\onlinecite{Alekhin:2016wb}, is plotted for majority (red) and minority (blue) carriers in the direction Fe $\to$ Au (dashed lines) and Au $\to$ Fe (solid lines).\label{Figure1}}
	\end{figure}
	
	In the following, we split the distribution function 
	\begin{equation}
	f_{\sigma}(\mathbf{r},\mathbf{k},t) = f_{\sigma}^{\mathrm{eq}}(\mathbf{r},\mathbf{k}) + g_{\sigma}(\mathbf{r},\mathbf{k},t)\, .
	\end{equation}
	into a contribution $f^{\mathrm{eq}}$ of equilibrium carriers and a contribution $g$ of hot electrons. In the linear regime at metallic densities, we have $g\ll f^{\mathrm{eq}}$. We assume $g_{\sigma}(\mathbf{k}) = 0$ for $|\mathbf{k}|<|\mathbf{k_{\mathrm{F}}}|$ and neglect contributions from holes, which propagate more slowly because of their band curvature.~\cite{Nenno:2016cm} Note that by treating only hot electrons, we neglect holes. 
	To arrive at a numerically tractable model, we first specialize the full transport equation and then use the particle-in-cell (PIC) approach for the numerical solution. The PIC approach leads to classical equations of motion for superparticles that represent the hot-carrier distribution function in a statistical fashion. We use the results of \textit{ab initio} calculations for real material data in our simulations. The quantities introduced in the following approximations and used as parameters are shown Fig.~\ref{Figure1}; they will be discussed later.
	
	We first split the dependence on $\mathbf{k}$ into its angular part, represented by the angle $\theta$ (with respect to the propagation axis $z$) and the energy $E$, which is connected to the wave vector through the band structure. Next, we follow closely the approximations made for the scattering term in Ref.~\onlinecite{Penn:1985wt}, however without taking the angular average of the distribution to describe transport in the multilayer.
	
We assume the scattering amplitude $w$ to be isotropic in $\mathbf{k}$. This so-called random-$k$ approximation~\cite{Penn:1985wt,Zhukov:2006ky} means that in-scattered particles have a completely random propagation direction, and has also been used recently in Ref.~\onlinecite{Battiato:2016bi}. The last term in Eq.~\eqref{eq:bte} can then be rewritten as
	\begin{equation}
	\begin{split}
	& \sum_{\sigma'}\int d\mathbf{k'}w(\sigma',\mathbf{k'};\sigma,\mathbf{k})g_{\sigma'}(\mathbf{k}',t) \\ = 
	&\sum_{\sigma'}\int\frac{d\Omega'}{4\pi}\int dE'\,w(\sigma',E';\sigma,E)g_{\sigma'}(\mathbf{r},E',\theta',t)\rho_{\sigma'}(E') \, ,
	\label{eq:scattering_term}
	\end{split}
	\end{equation}
where the integration over $d\Omega = d\varphi d\cos\theta$ removes any angular dependence and we neglected the spatial dependence of the scattering amplitude $w$ and the density of states of hot carriers, $\rho$, for simplicity. In replacing $f$ with $g$, we have neglected scattering events between hot carriers. We include only the effect of their scattering with equilibrium electrons on the hot-electron distribution. The scattering amplitude $w$ includes both inelastic and as elastic scattering events and we split $w$ according to
	\begin{equation}
	w(\sigma',E';\sigma,E) = w_{\mathrm{in}}(\sigma',E';\sigma,E) + w_{\mathrm{el}}\, .
	\label{eq:split_w}
	\end{equation}
where $w_{\mathrm{in}}$ denotes the inelastic scattering amplitude, $w_{\mathrm{el}}$ the elastic scattering amplitude. Elastic scattering events do not change the energy, nor the spin of the hot electron.
	%Secondary carriers are excited in every inelastic scattering event to fulfill the Pauli-principle in the original Coulomb-scattering term.~\cite{Penn:1985wt} 
In an inelastic Coulomb-scattering event, one hot electron scatters with equilibrium carriers to generate two carriers above the Fermi level. Ref.~\onlinecite{Penn:1985wt} shows that the following relation holds between out- and in-scattering contributions,
	\begin{equation}
	\frac{1}{\tau_{\sigma'}(E')} = \frac{1}{2}\sum_{\sigma}\int^{E'}_{E_{\mathrm{F}}} dE\, \rho_{\sigma'}(E') w_{\mathrm{in}}(E',\sigma';E,\sigma) \,,\label{eq:scat_time_equals}
	\end{equation}
where $\tau_{\sigma'}(E')$ denotes the inelastic lifetime, cf. Eq.~\eqref{eq:lifetimes}. This expression includes secondary carrier generation in the factor 2.
	We now assume that the inelastic in-scattering probability is independent of the final state $E$, in analogy to the derivation of the superdiffusive transport equations.~\cite{Battiato:2010br} Equation~\eqref{eq:scat_time_equals} can be rearranged to
	\begin{align}
	w_{\mathrm{in}}(E',\sigma';E,\sigma) = \frac{1}{E'-E_{\mathrm{F}}} \frac{2}{\rho_{\sigma}\tau_{\sigma'}(E')} \, ,
	\label{eq:inelastic_collision}
	\end{align}
	leading to an energy-dependent term in Eq.~\eqref{eq:bte} that distributes the carriers to all lower energies with equal probability, weighted by the relaxation time. 
	
Scattering effects due to impurities or phonons are considered to be elastic and thus only change the propagation angle and not the energy. The elastic scattering time in both iron and gold is assumed to be 30\,fs and the scattering process completely randomizes direction of propagation.~\cite{Kaltenborn:2014hg}
	This part of the scattering term can then be written as
	\begin{align}
		\begin{split}
			w_{\mathrm{el}} = \frac{\delta(E-E')}{\rho_{\sigma'}(E')\tau_{\mathrm{el}}} \, .
			\label{eq:elastic_scattering}
		\end{split}
	\end{align}
	The elastic scattering time $\tau_{\mathrm{el}}$ is assumed to be independent of energy.~\cite{Kaltenborn:2014hg} During elastic scattering events, spin is conserved.
	
As in Ref.~\onlinecite{Nenno:2016cm}, only the carrier transport perpendicular to the layers is modeled here. Typical laser spot sizes are on the order of several~$\mu$m, such that inhomogeneities in carrier transport in the plane of the layers are negligible on the timescales considered.~\cite{Shokeen:2017uu} The transport term in Eq.~\eqref{eq:bte} then reduces to $v_{\sigma}(E)\cos(\theta)\partial_z$, where again~$\theta$ is the propagation angle with respect to the axis of the layer stacking, $z$, and $v(E)$ is the energy-dependent magnitude of the propagation velocity obtained from \textit{ab initio} calculations.
	
Neglecting the influence of internal fields due to small charge accumulation and a short screening length~\cite{Zhu2014-oy} and in the absence of any external field yields the evolution equation for the hot-carrier distribution function, which will be used in the rest of the text, 
	\begin{widetext}
	\begin{equation}
	\begin{split}
	\bigg[\frac{\partial}{\partial t} + v_\sigma(E)\cos(\theta) \frac{\partial}{\partial z}\bigg] g_{\sigma}(z,E,\theta,t)= &S_{\sigma}(z,E,t) - \frac{g_{\sigma}(z,E,\theta,t)}{\tau^{\mathrm{eff}}_{\sigma}(z,E)} \\ 
	&+ 
	\sum_{\sigma'}\int\frac{d\Omega'}{4\pi}\int dE'\,w(z,\sigma',E';\sigma,E)g_{\sigma'}(z,E',\theta',t)\rho_{\sigma}(z,E) \, .
	\label{eq:bte2}
	\end{split}
	\end{equation}
	\end{widetext}
We solve this integro-differential equation using numerical methods described in the next section.
	
	%%%%%%%%%%%%%%%%%%%%%%
	% Numerical Approach %
	%%%%%%%%%%%%%%%%%%%%%%
	\section{Numerical Approach\label{sec:numerics}}	
	\subsection{Particle-In-Cell and Operator Splitting}
	Equation~\eqref{eq:bte} with the previously discussed approximations is solved using the PIC method combined with an operator-splitting technique. The operator splitting breaks the full problem into subproblems that are more easily solved numerically. We then sample the electron distribution function with superparticles that are found to obey classical equations of motion.~\cite{Faghihi:2017vv}
	
	For the PIC method, the distribution function is represented by a sum over all superparticles,	\begin{equation}
	\begin{split}
	g_{\sigma}(z,E,\theta,t) \simeq  C \sum_{i=1}^{N(t)} w_i \delta(z-z_i)\delta(E-E_i)\\ \times \delta(\theta -\theta_i)\delta_{\sigma \sigma_i} \, .
	\end{split}
	\label{eq:PIC}
	\end{equation}
	The energies $E_i$, propagation angles $\theta_i$ and positions $z_i$ characterize the state of superparticle $i$ and are time-dependent. The superparticles are weighted by a prefactor $w_i$ in the sum~\eqref{eq:PIC}, which is adjusted to improve the representation of less likely events. The constant  $C=\rho_{\mathrm{exc}}/\sum_{i=1}^{N(t)} w_i$ in front of the sum~\eqref{eq:PIC} is a normalization to reproduce the excited carrier density correctly. The number $N(t)$ of superparticles can change during the dynamics when additional (secondary) electrons are created.
	
	Before we derive the equations of motion for the superparticles, we first apply the Strang operator splitting~\cite{Strang:1968dr} to Eq.~\eqref{eq:bte2}. The operator splitting technique reduces the full problem,
	\begin{equation}
	\frac{\partial}{\partial t}g = \mathcal{L}_{\mathrm{Transport}}[g]	+ \mathcal{L}_{\mathrm{Source}}[g] 
	+\mathcal{L}_{\mathrm{Scattering}}[g] \, ,
	\label{eq:bte3}
	\end{equation}
	to simpler subproblems.  Every term on the RHS in Eq.~\eqref{eq:bte3} then leads to an individual partial differential equation (PDE) which is solved using the PIC approach. We choose an operator-splitting method for which all individual equations except for one are solved twice, each for half a timestep in a particular order to maintain second-order accuracy in time. Since the scattering part is numerically more demanding than the other terms, it is solved for the full timestep. More explicitly, we have
	\begin{align}
		\partial_t g_{\sigma} &= \mathcal{L}_{\mathrm{Transport}}[g_{\sigma}] = v_{\sigma}(E)\cos(\theta)\partial_z g_{\sigma} \, 
		\label{eq:Strang-transport},\\
		\partial_t g_{\sigma} &= \mathcal{L}_{\mathrm{Source}}[g_{\sigma}] = S_{\sigma}(z,E,t)\, , \label{eq:Strang-source}
	\end{align}
	and
	\begin{equation}
	\begin{split}
	\partial_t g_{\sigma} = \,&\mathcal{L}_{\mathrm{Scattering}}[g_{\sigma}]= - \frac{g_{\sigma}}{\tau^{\mathrm{eff}}_{\sigma}(E)} \\&+ \sum_{\sigma'}\int d^3k'w(\sigma',E';\sigma,E)g_{\sigma'}(z,E',\theta',t)\rho_{\sigma}(z',E) \, .
	\end{split}
	\label{eq:Strang-scattering}
	\end{equation}
	The solution to Eq.~\eqref{eq:bte3} is then the concatenation of the individual operators applied to the initial distribution,
	\begin{equation}
	\begin{split}
	g^{(1)} &= \mathcal{L}_{\mathrm{Source}}(\Delta t/2)[g(t)] \, ,\\
	g^{(2)} & = \mathcal{L}_{\mathrm{Transport}}(\Delta t/2)[g^{(1)}] \, ,\\
	g^{(3)} & = \mathcal{L}_{\mathrm{Scattering}}(\Delta t)[g^{(2)}]\, , \\
	g^{(4)} & = \mathcal{L}_{\mathrm{Transport}}(\Delta t / 2)[g^{(3)}]\, , \\
	g(t+\Delta t) & = \mathcal{L}_{\mathrm{Source}}(\Delta t/2)[g^{(4)}] \, .
	\label{eq:strang}
	\end{split}
	\end{equation}
	All three operators need to be numerically solved with second-order accuracy in time for the full five-step solution to be accurate to second order as well.~\cite{Strang:1968dr} We now apply the PIC representation~\eqref{eq:PIC} for $g$ to Eq.~\eqref{eq:bte3} and calculate the macroscopic moments with respect to $g$,
	\begin{align}
		\langle X \rangle_{g_{\sigma}} &= \sum_{\sigma}\int dE\int\frac{d\Omega}{4\pi}\int dz \,X\,g_{\sigma}(z,E,\theta,t) \, ,\label{eq:firstmoment}
	\end{align} 
	where $X=1,z,E,\theta,\sigma$.
	This procedure yields classical equations of motion for the quantities $N_i,\,z_i,\,E_i,\,\theta_i$ and $\sigma_i$ that characterize the superparticles and which are determined by different processes described by the full Boltzmann transport equation~\eqref{eq:bte3}. 
	
	% Figure 2
	\begin{figure}[t]
		\includegraphics[width=0.45\textwidth]{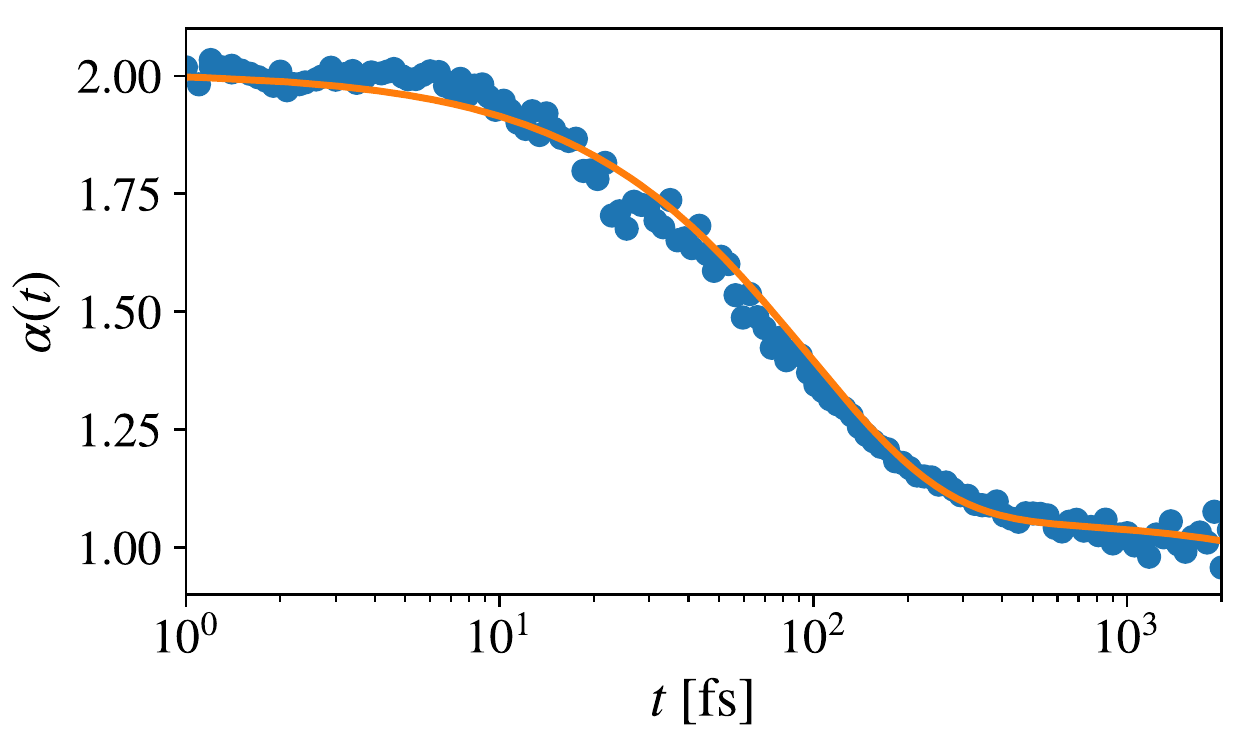}
		\caption{Numerical (dots) and analytical result (line) for the generalized diffusion coefficient over time for a mono-energetic excitation in an infinite slab with only elastic scattering included ($\tau_{\mathrm{el}}=30\,\mathrm{fs}$).\label{Figure2}}
	\end{figure}
	
	% Number of Particles
	First, we find an equation for the number of superparticles by calculating $\partial _t \langle 1 \rangle$, which results in 
	\begin{equation}
	\partial_t N(t) = \sum_i S_{\sigma_i}(z_i,E_i,t) \, .
	\end{equation}
	Superparticles are activated by this source term and numerically, they are created in such a way that they yield a good sampling of $g$. Their total number is a numerical parameter that needs to be checked for convergence. 
	We choose a Gaussian pulse shape in time and excite particles homogeneously in the excited part of the structures under consideration and at fixed energy. Since the excitation process does not favor any propagation direction, the propagation angle is randomly chosen. We note, however, that due to the finite size of the structures and the particle energy below the ionization potential, carriers can only travel into the interior of the material. At the beginning and end of every timestep, according to Eq.~\eqref{eq:strang}, we adjust the particle number depending on~$S$.
	
	% Transport part
	For the particle position, we find
	\begin{equation}
	\frac{\partial z_i(t)}{\partial t}= v_{\sigma_i}(E_i)\cos(\theta_i) \,,
	\end{equation}
	a classical equation of motion that is easy to solve. Neither energy nor the propagation angle changes during this process. As this equation results from the transport part, it is solved twice for each half a timestep in the Strang splitting~\eqref{eq:strang}. In addition, for carriers that would cross the vacuum interface, we include a full reflection back into the material. At the interface between two materials, the probability that the superparticle moves through the interface is treated as a stochastic process with a uniform distribution chosen to reflect the transmittance of the interface, $T_{\mathrm{Fe}\rightleftarrows	\mathrm{Au}}(E)$ from Fig.~\ref{Figure1}.
	
	% Scattering part
	The scattering contribution~\eqref{eq:Strang-scattering} is treated with a Monte-Carlo approach. First, since Eq.~\eqref{eq:bte2} includes both inelastic and elastic events, we calculate the probability for one of the events to happen within $\Delta t$, comparing the timestep to the effective scattering lifetime $\tau^{\mathrm{eff}}_{\sigma}$. The timestep is a numerical parameter that is kept constant during the calculation and thus becomes a quantity that we have to converge, even though the solution of the transport problem, for example, does not depend on it. 
	
	Secondary carrier generation is included in this approach and is implemented in such a way that secondary carriers are generated with the equilibrium polarization 
	\begin{equation}
	P_0(z) =  \frac{n^0_{\uparrow}(z) - n^0_{\downarrow}(z)}{n^0_{\uparrow}(z) + n^0_{\downarrow}(z)} \, .
	\label{eq:eq_pol}
	\end{equation}
	Here, $n^0_{\sigma}$ denotes the equilibrium carrier density in each spin channel. This quantity vanishes in the non-magnetic part of the slab.	
	
	For \emph{elastic scattering}, the particle number is conserved. The integral in Eq.~\eqref{eq:scattering_term} is then solved stochastically by randomizing the propagation angle of the superparticle, while keeping the energy constant. In case of an \emph{inelastic collision}, a final energy for the scattering superparticle is chosen uniformly, according to Eq.~\eqref{eq:inelastic_collision}. If such a collision occurs, a superparticle from below the Fermi energy is elevated to an energy that is smaller than the final energy state of the initial carrier. Thus energy is conserved in the scattering, however the hot electron energy can decrease. With this approach, we solve both the in- and out-scattering (relaxation) term in the BTE~\eqref{eq:bte2}. In the lowest energy bin considered, we remove particles from the simulation if these undergo an additional scattering event. This corresponds to these particles becoming quasi-thermalized so that they no longer contribute to the hot-electron dynamics.
	
	%%%%%%%%%%%%%%%%%
	% Model Studies %
	%%%%%%%%%%%%%%%%%
	\section{Model Studies\label{sec:model_studies}}
	% Figure 3 
	\begin{figure}[t]
		\includegraphics[width=0.45\textwidth]{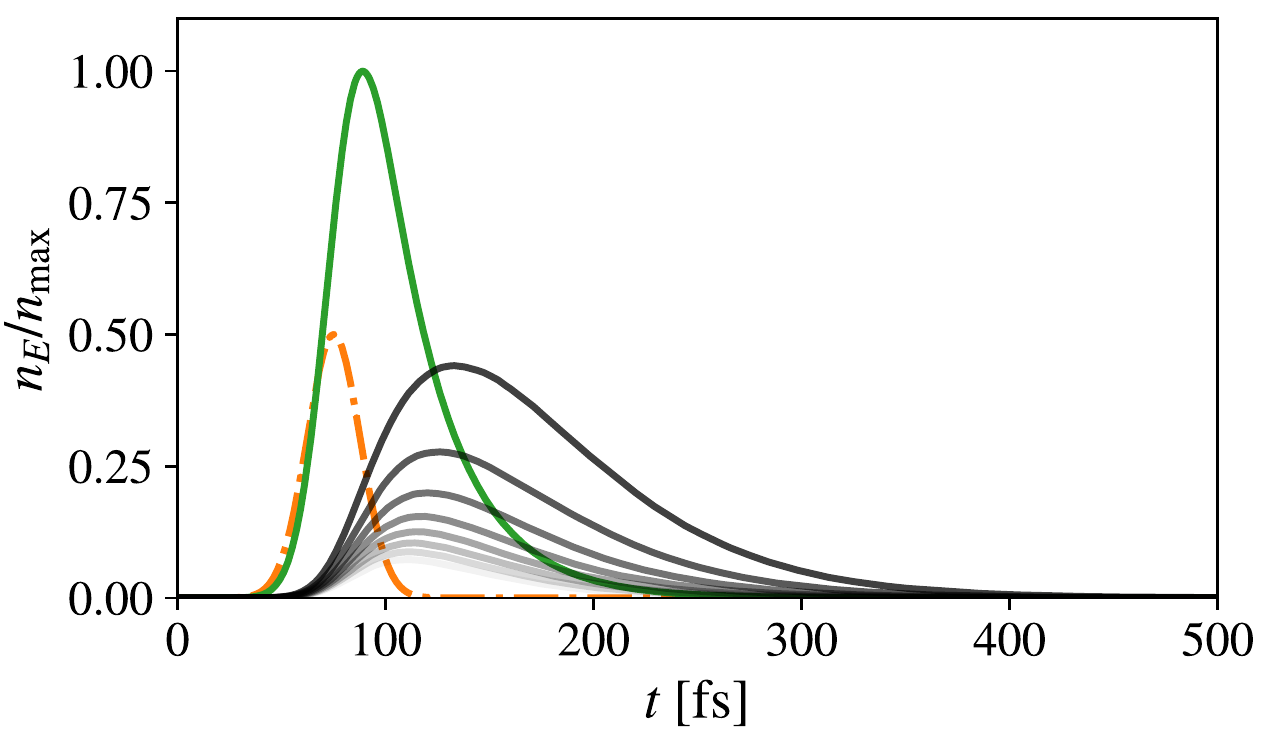} 
		\caption{Density over time for carriers of discrete energy (darker for lower energies, green line represents carriers with the excitation energy). The relaxation time is kept constant ($\tau_{\mathrm{in}}(E)=30\,\mathrm{fs}$) over the range of energies. The duration and strength of the laser pulse is indicated by the dashed line.
			\label{Figure3}}
	\end{figure}
	We first investigate the general behavior of the transport of excited carriers and how it depends on the specific choice of parameters. Due to the nature of our model, both the spatial transport and the source term are solved analytically and it is the number of particles and the dynamics due to the scattering term that require checking for convergence. 
	
	Typically, we use up to 10 million first-generation particles that initiate the dynamics and excite more carriers over time. Our results for simulations up to 1 picosecond do not improve above this order of magnitude.
	
	We first discuss the generalized diffusion coefficient
	\begin{equation}
	\alpha = t \frac{1}{\sigma^2} \frac{d\sigma^2(t)}{dt}\,,
	\end{equation}
	which is connected to the the mean-square displacement $\sigma^2(t) =  \langle z^2\rangle_g$ obtained from the dynamical distribution $g$ by the relation~\cite{Metzler:2000hea} $\sigma^2(t)\propto t^{\alpha}$. The generalized diffusion coefficient is used to characterize the motion of an ensemble of carriers, whose characteristics vary between completely ballistic ($\alpha = 2$) and purely diffusive ($\alpha = 1$) behavior. Fig.~\ref{Figure2}, shows this generalized diffusion coefficient computed using our numerical approach for the case of an infinite slab with a mono-energetic excitation at $E=1.5$\,eV and \emph{only} elastic scattering effects ($\tau_{\mathrm{el}}=30\,\mathrm{fs}$) included. Compared to the analytical result for this case,~\cite{Weiss:2002bb}
	\begin{equation}
	\alpha(t) = \frac{t (1 - e^{-t/\tau})}{t - \tau( 1- e^{-t/\tau})}\, ,
	\end{equation}
	we find very good agreement, with small disparities coming mainly from numerically extracting the exponent from the computed mean-square displacement.
	
	We find the same characteristic transition from ballistic behavior ($\alpha=2$) at short times to a diffusive behavior ($\alpha=1$) at longer times that was obtained in Ref.~\onlinecite{Battiato:2012hw} using a half-analytical calculation.  Over several hundred femtoseconds the transport behavior can be classified as ``superdiffusive''. Note that this superdiffusive behavior in the transition region occurs without inelastic scattering of the excited carriers nor creation of secondary electrons. In accordance with Ref.~\onlinecite{Battiato:2012hw}, we stress that for typical sub-picosecond experimental time scales neither a fully ballistic nor a completely diffusive ansatz well describes the carrier dynamics. 
	
	Figure~\ref{Figure3} shows results obtained for an infinite layer including inelastic scattering processes. We plot the evolution of the particle density in different energy intervals (``bins'') of 75 meV size around the indicated energies. The excitation is modeled as a Gaussian laser pulse that injects electrons in the bin with the highest energy, which leads to an increase of particle density in that bin. The particle density in the highest-energy bin subsequently decreases due to out-scattering events which increases the density of electrons in bins at lower energies. The particle density in the bins at high energies peak earlier and at a lower maximum value than those at lower energies. This reflects the down scattering of excited electrons and production of secondary electrons which accumulate in the lower-energy bins. 
	
	Due to the changing number of particles in our simulation, it is numerically difficult to extract the diffusion coefficient if we include inelastic scattering events, but we expect a qualitatively similar result to Fig.~\ref{Figure2}, with an effective scattering time which now includes inelastic effects. As long as the inelastic scattering time is longer than the elastic scattering time, the resulting generalized diffusion coefficients should be similar, even though the underlying electron dynamics are different.
	
	%%%%%%%%%%%%%%%%%%%%%%
	% Simulation Details %
	%%%%%%%%%%%%%%%%%%%%%%
	\section{Simulation Results\label{sec:results}}

	We now apply the approach described in the previous sections to simulate the excited-electron dynamics in different multilayer structures composed of gold and iron. We solve Eq.~\eqref{eq:bte2} numerically using the \textit{ab initio} data for these materials shown in Fig.~\ref{Figure1}. Carrier velocities are taken from Ref.~\onlinecite{Zhukov:2006ky}, lifetimes of majority and minority carriers in iron from Ref.~\onlinecite{Kaltenborn:2014du}. The velocities of hot carriers in gold were calculated using the free electron model, the energy-dependent lifetimes are modeled by a Fermi-liquid fit to experimental data.~\cite{Cao:1998ec} Transmission and reflection probabilities for a Fe/Au-interface are taken from Ref.~\onlinecite{Alekhin:2016wb}. 
	
	In Fig.~\ref{Figure1} we also show the mean-free paths derived from lifetimes and carrier velocities, which play an important role in the interpretation of the calculated results below. Optical fields excite electrons to energies of about 1 to $1.5$\,eV where the spin-dependent velocities are different: majority velocities are between $0.5$ and 1\, nm/fs, whereas minority velocities are below 0.2\,nm/fs. Since the lifetimes in this energy range are similar, the mean-free paths reflect the difference of an order of magnitude between the velocities. Despite this difference in mean-free paths, however, the electronic dynamics are neither completely ballistic nor diffusive for times of a few hundred femtoseconds. 
	
	This is taken into account in our calculations by the energy, space and time-dependent distribution functions, which are shown in a position integrated way in Fig.~\ref{Figure3}. It is the rather complicated spatio-temporal dynamics that underlie the integrated polarization profiles that we will discuss in this section, and it is an oversimplification to model these dynamics using the mean-free path as a length scale that indicates a sharp transition between ballistic and diffusive transport.~\cite{Kaltenborn:2014hg,Nenno:2016cm}
	
	\subsection{Fe-Au Bilayer\label{sec:bilayer}}
	%%%%%%%%%%%%%%%%%%%%%%
	% Simulation Results %
	%%%%%%%%%%%%%%%%%%%%%%
	% Long slabs
	We first analyze the case of a bilayer system composed of 15\,nm iron and 100\,nm gold. The excitation is modeled after a 30\,fs infrared laser pulse with a photon energy of~$1.5$\,eV, which elevates majority carriers in iron to~1.5\,eV above the Fermi level and minority carriers to~$E_{\mathrm{F}}+1$\,eV.~\cite{Melnikov:2011epa,Nenno:2016cm} From the computed time and space dependent majority and minority carrier densities we obtain the spatio-temporal profile of the polarization of excited electrons,
	\begin{equation}
	P_{\mathrm{hot}}(z,t) = \frac{n^{\mathrm{hot}}_{\uparrow}(z,t) - n^{\mathrm{hot}}_{\downarrow}(z,t)}{n^{\mathrm{hot}}_{\uparrow}(z,t) + n^{\mathrm{hot}}_{\downarrow}(z,t)} \, .
	\label{eq:hot_carrier_polarization}
	\end{equation}
	We designate the density of the simulated electrons by~$n^{\mathrm{hot}}$ to stress that we track only electrons with energies more than 0.1 eV above the Fermi sea, and we do not take into account the formation of a quasi-equilibrium. The polarization is normalized to the total number of hot carriers and thus, in the linear regime considered here, does not depend on the excitation strength. We also investigate the spin-current density $j_s$ carried by the hot electrons, which we obtain from
	\begin{equation}
	j_s(z,t) = -e \big[ \langle v_{\uparrow}\rangle_{g_{\uparrow}} - \langle v_{\downarrow}\rangle_{g_{\downarrow}} \big]\, .
	\label{eq:spincurrent}
	\end{equation}
	The magnitude of the spin-current density is proportional to the density/number of excited hot-carriers and we therefore use arbitrary units in the following. Note that the polarization~\eqref{eq:hot_carrier_polarization} captures certain aspects of the hot-electron transport, but it is a normalized quantity that can show changes due to small absolute numbers of electrons. Here, the spin current yields useful additional information on the spatio-temporal transport, as it depends on the number density of transported electrons. Typical excitation densities are on the order of 0.1 electrons per atom in the emitter layer~\cite{PhysRevLett.85.844,Battiato:2012hw}, resulting in induced hot-carrier spin currents into the non-magnetic material on the order of $10^{29} \hbar/(\mathrm{m}^2\mathrm{s})$. We drop the units in the following discussion, but will discuss their relative strengths.
			
	%%%%%%%%%%%
	% Bilayer %
	%%%%%%%%%%%
	For the bilayer structure, and also for the spin-valve structure discussed below, we compare the transport dynamics for three different cases: (1) the ``full'' calculation including secondary-electron generation and interface reflectance/transmittance, (2) a calculation including secondary-electron generation but a transparent interface to highlight the importance of the interface, and (3) a calculation that includes the reflectance/transmittance of the interface, but not the generation of secondary electrons.

	% Figure 4 - Bilayer Fe/Au
	\begin{figure*}[t]
		\includegraphics[width=0.975\textwidth]{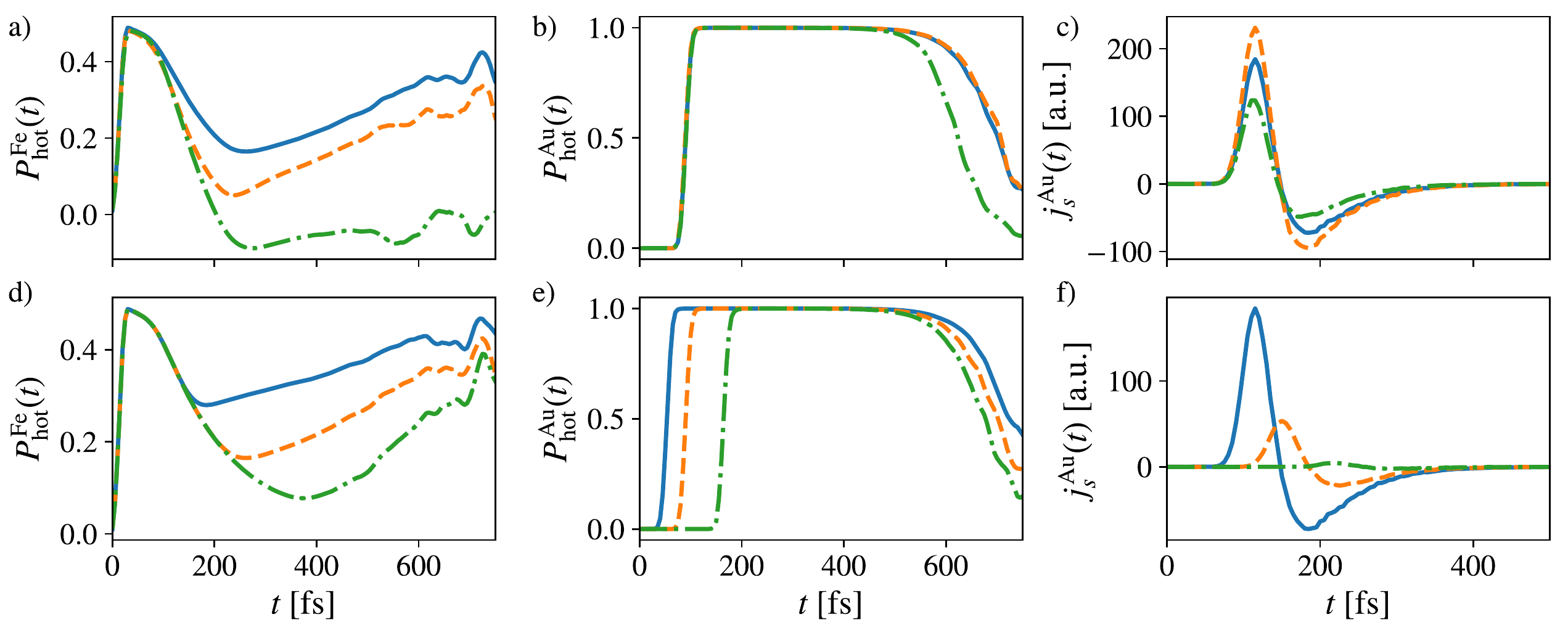}
		\caption{Computed results comparing different simulations for the Fe-Au bilayer (a)--(c): full calculation (solid), transparent interface (dashed), no secondary electrons (dash-dotted). For the Fe (15nm)-Au(100nm) system, subplots (a) and (b) show the hot-carrier polarizations vs.~time averaged in the Fe layer and 5nm from the gold-vacuum interface, and (c) the spin current at the vacuum interface of the Au layer. Corresponding results computed with the full simulation for thicknesses of 50nm (solid), 100nm (dashed) and 200nm (dash-dotted) of the Au layer (d)--(f).\label{Figure4}}
	\end{figure*}

	Figure~\ref{Figure4} shows the time evolution of the hot-electron spin polarization in the layers and the corresponding spin current at the Fe-Au interface. We first discuss the polarization in the Fe layer shown for the three different simulations in Fig.~\ref{Figure4}(a) and an excitation by a laser pulse of 30 fs centered at $t=75$ fs. In all three cases, the polarization rises due to the dominant generation of primary majority electrons. It does not exceed 50\%, which is close to the equilibrium polarization of Fe d-electrons.~\cite{Soulen85} and drops immediately after reaching its maximum value. This behavior arises because majority carriers in iron have higher velocities and transmission probabilities, as shown in Fig.~\ref{Figure1}, and can leave the slab immediately after they have been excited. After the laser pulse is ramped down, the polarization is reduced as the hot electrons leave the slab. This is a key point for the interpretation of the computed hot-electron polarization. We assume that the hot primaries are generated by the optical field with a polarization equal to that of the equilibrium polarization. When the excited carriers undergo transport ans scattering processes, the hot-electron polarization is not directly related to the total carrier polarization anymore, as we only keep track of electrons above the Fermi level. Around 100\,fs, most primary majority carriers leave the slab such that the hot-electron polarization in the Fe layer is reduced. In the case without secondary carrier generation the hot-electon polarization even drops below zero. As mentioned before, only for large enough densities and at short times, this reduction of hot-electron polarization corresponds to a demagnetization of the slab as majorities leave the slab more quickly than they relax back to the Fermi energy within the region where they were excited. Minority carriers have a low transmission probability, so that they are more likely to stay within their excitation region. Accordingly, for a transparent interface, the hot-electron polarization drops faster than in the full calculation but the hot electrons do not completely depolarize because in this case secondary carrier generation produces additional carriers with a polarization equal to the equilibrium polarization of the ferromagnet. After about 250\,fs the polarization rises again when electrons that have propagated through the whole structure are reflected and reenter their excitation region. The fluctuations at longer times occur when the number of hot carriers strongly reduced so that the statistics become worse, even for numerical calculations using one billion primary particles. We now turn to the corresponding hot-electron dynamics in the Au layer. It shows a delay due to the initially ballistic propagation through this Au layer, which results in positive hot-electron polarization due to incoming ballistic Fe-majority carriers. Fe-minority electrons propagate through the Fe slab more slowly and have a low transmission probability, such that their influence in gold is weak. The hot-electron polarization goes to zero  after approximately 0.5\,ps, and the influence of secondary carrier generation is clearly seen as it produces a slightly longer positive signal due to effectively slower electrons.
	
	In Fig.~\ref{Figure4}(c) we plot the spin-current flowing in the Au layer close to the vacuum interface at the right side of the structure. The current dynamics shows a strong ``bipolar'' signal and essentially vanishes after 250\,fs. This behavior at longer times shows that the hot-electron polarizations visible after 250\,fs in Fig.~\ref{Figure4}(a),(b) are not significant as they do not correspond to the flow of a spin current. Instead, they originate from a small number of carriers, as the spin polarization is normalized by the total number of excited carriers and calculated by Eq.~\eqref{eq:hot_carrier_polarization}. The bipolar spin-current signal arises from Fe-majority (``$\uparrow$'') carriers transported into the gold and the subsequent reversal of the current mainly by reflected $\uparrow$ electrons and also by slower $\downarrow$ electrons. The height of the current signal corresponds to the rise and drop of the polarization in the Fe layer: Including only primary carriers results in the weakest spin-current, a transparent interface through which all excited electrons can escape the ferromagnet results in the strongest. The full calculation lies in between.
	
	Figures~\ref{Figure4}(d)--(f) show the results of the full calculation for an Fe(15nm)-Au(x) bilayer for different thicknesses $x=50$\,nm, 100\,nm and 200\,nm of the Au layer. The hot-electron polarizations in Fe and Au, shown in Fig.~\ref{Figure4}(d) and (e), respectively, show a similar behavior to the one just discussed. The Fe polarization drops and rises again when majority electrons first leave that the Fe slab and come back after traversing the entire structure. Again, the transmission from gold into iron is high for majority carriers as well, such that they propagate almost freely throughout both layers. This is reflected by the rising polarization in Au due to incoming ballistic majority carriers in all cases. Increasing the Au layer thickness shifts the curve by the ballistic propagation time of the fastest carriers in Au and slightly shortens the duration of the positive plateau due to more intermediate scattering events. Finally, the spin current in Fig.~\ref{Figure4}(f) shows a weakening bipolar signal with increasing thickness of the Au layer; for the 200\,nm layer the signal almost vanishes as the density of hot carriers is lowered as electrons are scattered to lower energies on the length scale of the inelastic mean free path. These low-energy electrons are not included in out calculation. In addition, the signal becomes shorter for increasing layer thickness. The later onset in thicker layers, again, is due to ballistic primaries. The shorter length of the signal however is due to the fact that for thicknesses over 50 nm, the mean-free path of the fastest carriers is shorter than the layer thickness. Thus, those carriers can relax back to the Fermi level and are not considered in the hot-carrier spin current.
	
\begin{figure}[t]
		\includegraphics[width=0.975\columnwidth]{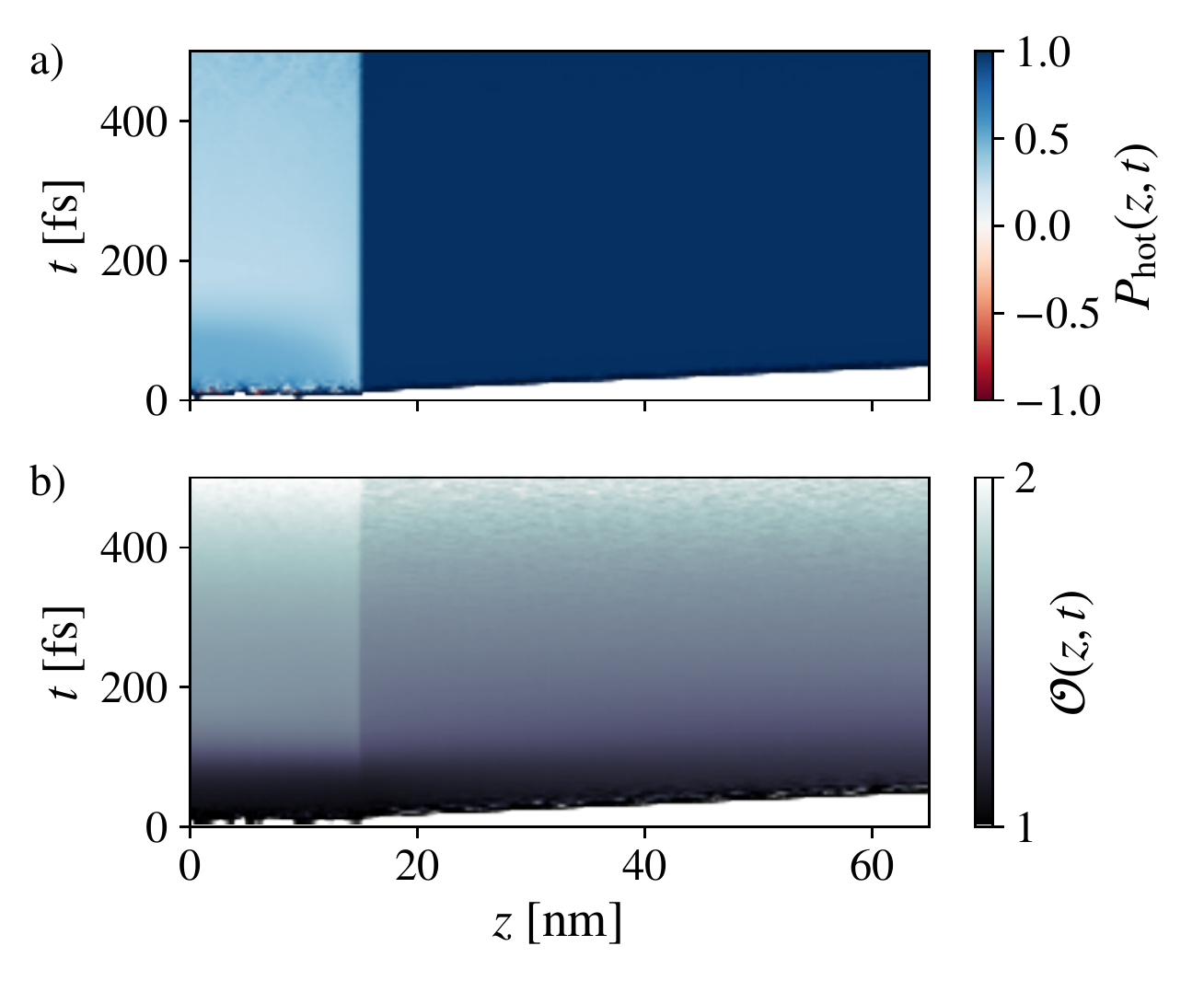}
		\caption{Computed space and time resolved hot-carrier polarization $P_{\mathrm{hot}}$ (a) and generation of carriers (b), $\mathcal{O}$, for the Fe(15nm)-Au(50nm) structure. Each discrete superparticle is assigned a generation label (1st: primary, 2nd: secondary carriers, etc.), which is averaged in each phase space cell; the space-time map of this generation label is shown in (b). \label{Figure5}}
	\end{figure}

	To investigate this behavior in more depth, we plot the hot-carrier polarization over time and space for a Fe(15nm)/Au(50nm) heterostructure in Fig.~\ref{Figure5}(a). Note the white wedge-shaped region at short times occurs because none of the excited carriers have reached these space-time points yet. In the iron slab the initial excitation leads to a homogenous polarization, which then slowly and homogeneously decays. The polarization in Au is dominated by majority carriers injected from the Fe layer and only relaxes relatively slowly. Fig.~\ref{Figure5}(b) contains further details on the scattering behavior. We track the particle generation and plot its mean value $\mathcal{O}(z,t)$ in every space-time cell. A value of $\mathcal{O}$ is assigned to primary carriers, whereas secondary carriers have generations $\mathcal{O}>2$. By plotting the generation of carriers in the iron layer, we find that primary carriers quickly scatter and excite secondaries and electrons with higher generations. Inelastic scattering in gold happens at a slower rate and higher electron generations occur only at later times, since the inelastic scattering time in gold is longer than the one in iron, cf.~Fig.~\ref{Figure1}(b). 
	However, that a clear distinction between ballistic and diffusive carriers in these structures is misleading, since the generation of secondaries that propagate ballistically and the inhibited transport due to interface scattering alters the transport behavior of the initially free hot-electron population. Fig.~\ref{Figure5}(b) can provide some insight here by showing that secondaries are fewer and have far longer reach in gold than in iron.

	% Figure 5 - Parallel and Antiparallel Trilayer

	%%%%%%%%%%%%
	% Trilayer %
	%%%%%%%%%%%%
	\subsection{Fe-Au-Fe Spin-Valve Structure}
	
	% Figure 6
	\begin{figure*}[t!]
		\includegraphics[width=\textwidth]{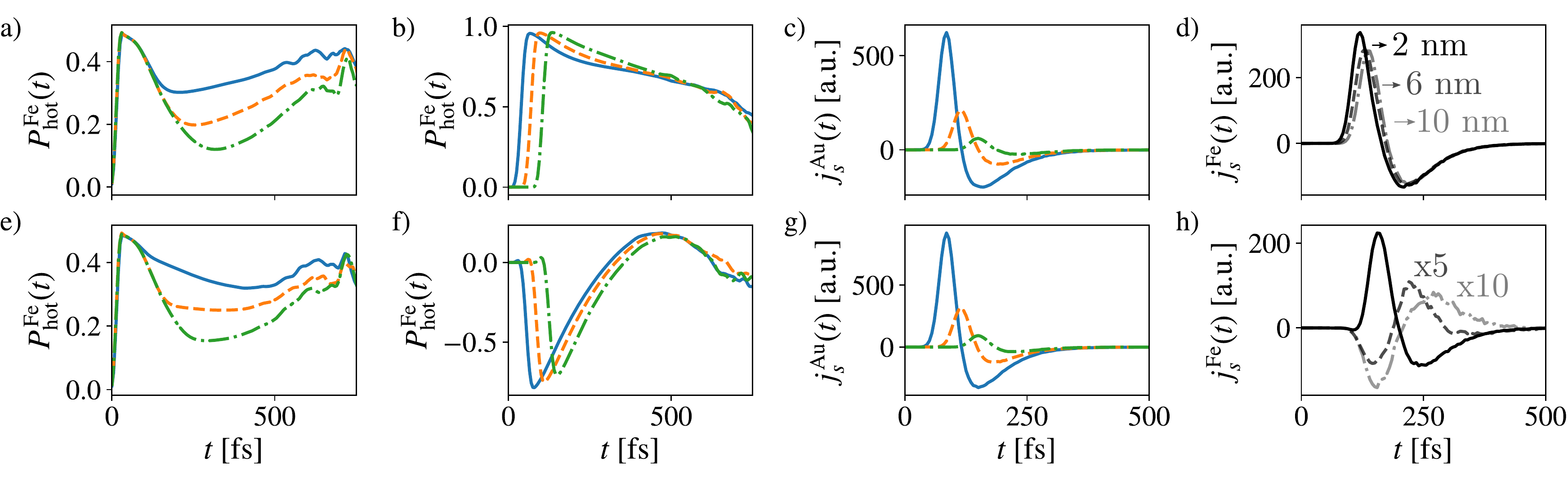}
		\caption{Computed results for the Fe-Au-Fe spin-valve structure with different collinear alignment of emitter and collector layers. $\uparrow$-$\uparrow$ results are shown in (a)--(d), and $\uparrow$-$\downarrow$ results in (e)--(h). Different Au layer thicknesses are color-coded: 10nm (solid blue), 50nm (dashed orange) and 100 nm (dash-dotted green). Shown are the hot-carrier polarization in the optically excited emitter (a),(e) and collector (b),(f) Fe layers. Note that the polarization in positive $z$ direction (alignment of the emitter) is always counted as positive, so that a negative hot-electron polarization in the $\downarrow$ collector actually enhances the polarization of the collector. Subplots (c) and (g) show the spin currents injected into the Au layer at the Au-Fe interface. Panels (d) and (h) show the spin currents injected into the collector layer after traversing a 50 nm Au spacer layer at different depths of the collector (indicated in (d)). Note that for the antiparallel configuration, the spin currents at different depths have been multiplied with the indicated factors (h). \label{Figure6}}
	\end{figure*}
	Next, we study the case of different Fe(15nm)/Au(x)/Fe(15nm) trilayers composed of Fe layers capping a gold layer of different thickness  $x=10$\,nm, 50\,nm and 100\,nm. Such spin-valve structures have been studied extensively for close-to-equilibrium transport, but recently Alekhin \emph{et al.}~\cite{Alekhin:2016wb} demonstrated that excitation of these structures by ultrafast optical pulses leads to a non-thermal spin-Seebeck effect. We follow Ref.~\onlinecite{Alekhin:2016wb} and call the left Fe layer, which we assume is excited optically, the emitter and the right Fe layer the collector. Using our approach, we study the hot-electron transport for collinear parallel ($\uparrow$-$\uparrow$) and antiparallel ($\uparrow$-$\downarrow$) configurations of the ferromagnetic emitter and collector layers. Our approach allows us to go from the bilayer to the trilayer system of different thicknesses using essentially the same numerical setup, without having to introduce additional particle fluxes as in Ref.~\onlinecite{Battiato:2014kz}. We choose a fixed Fe layer thickness of 15\,nm because it leads to an almost complete absorption of the laser pulse within the emitter layer,\footnote{We have checked this using standard optical-transfer-matrix calculations.} so that we do not need to include the direct optical excitation of carriers in the Au layer or the collector. All other input data remain unchanged with respect to the bilayer simulation to allow for easier comparison.

	Figure~\ref{Figure6} contains the computed hot-carrier spin polarization of the front and back iron layers, as well as the profiles of the injected spin-current at the gold-iron interface and at different positions in the collector for a 50\,nm Au spacer layer. The results in Fig.~\ref{Figure6}(a)--(d) are for the parallel configuration, in Fig.~\ref{Figure6}(e)-(h) for the antiparallel configuration.  
	
\begin{figure*}[t!]
		\includegraphics[width=0.975\textwidth]{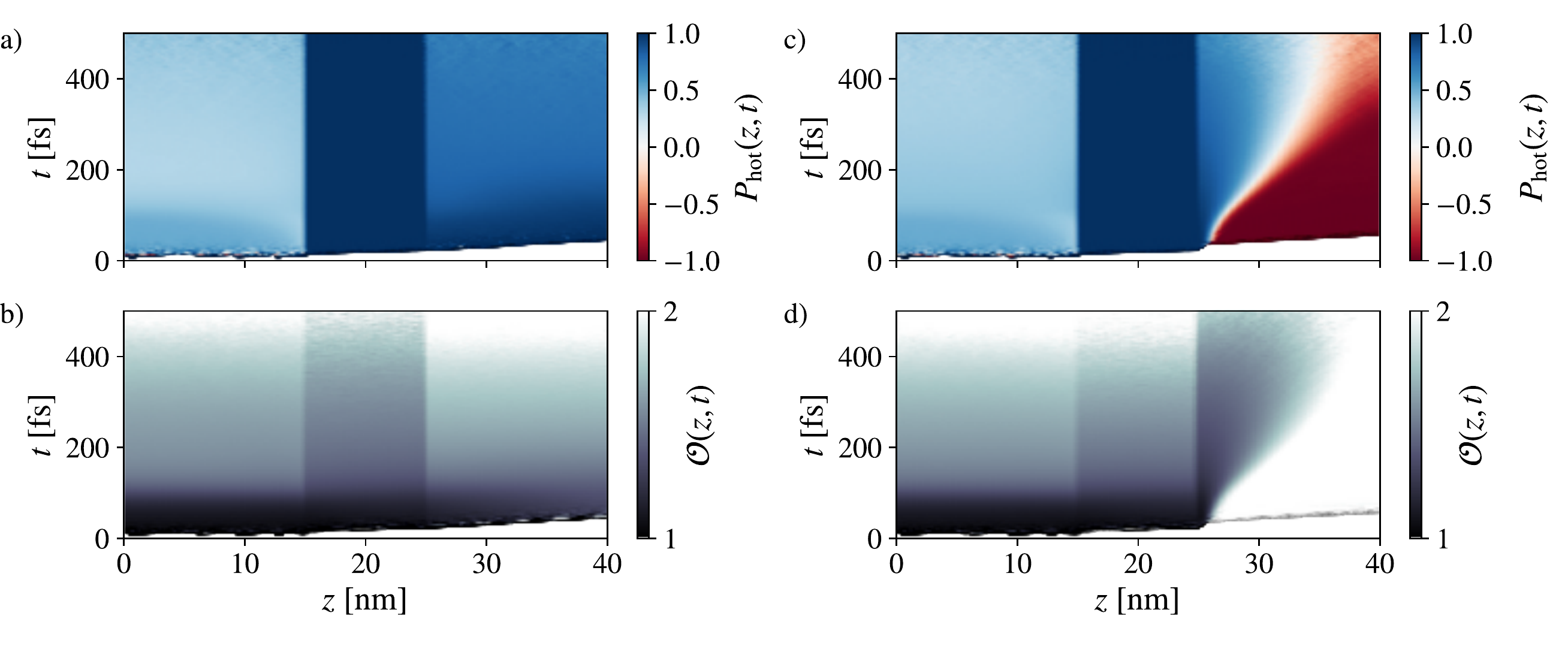}
		\caption{Computed space and time resolved hot-electron polarization (a),(c) and generation of carriers (b),(d) for the Fe(15nm)-Au(10nm)-Fe(15nm) spin-valve structure with different collinear alignment of emitter and collector layers. $\uparrow$-Au-$\uparrow$ results are shown in (a) and (b), and $\uparrow$-Au-$\downarrow$ results in (c) and (d). Each discrete superparticle is assigned a generation label (1st: primary, 2nd: secondary carriers, etc.), and the particles are averaged in each phase space cell, leading to non-integer results for the mean generation maps shown in (b) and (d). \label{Figure7}}
	\end{figure*}

	In both cases the emitter (left Fe layer) is optically pumped, and we show the hot-carrier polarization averaged in the ferromagnetic layers, as the optical pump excites the whole layer. In Fig.~\ref{Figure6}(a), which plots the polarization for the emitter in the parallel configuration, we observe a fast rise of the signal followed by drop after the pump pulse as observed in the bilayer structure. In analogy to the results discussed in Sec.~\ref{sec:bilayer}, reflected carriers at the Fe or vacuum interface then lead to an increase of polarization in the collector at later times. This positive tail of the signal varies in magnitude depending on the Au layer thickness and is stronger for short layers. Turning to the polarization profile of the collector shown in Fig.~\ref{Figure6}(b), we find a steep increase when majority carriers that were excited in the emitter enter the slab. In the parallel configuration, those majority carriers are almost completely polarized and enhance the polarization above the equilibrium value in the collector layer, due to the favorable transmission at the Au-Fe interface. They also act via scattering processes as an additional hot-electron generation process. Both initial and excited carriers can leave the Fe layer easily, such that the signal decreases after approximately 200\,fs. Fig.~\ref{Figure6}(c) shows the injected spin-current at the Au-Fe interface, which exhibits a similar behavior as seen in the bilayer structure. The spin-current in the collector of the structure with a 50\,nm Au spacer is shown in Fig.~\ref{Figure6}(d) at positions 2\,nm, 6\,nm and 10\,nm into the layer. We find a decrease in magnitude compared to the spin current in Au because of the influence of the interface. With increasing penetration depth the magnitude of the spin current slowly decreases as well, while the temporal profile remains almost unchanged. This suggests a slow decrease of the initial carrier population in the third iron layer and a spatially homogeneous spin-polarization. We will further discuss this using Figure~\ref{Figure7}(a).
	
	% Figure 7

	We now turn to the antiparallel ($\uparrow$-$\downarrow$) spin-valve configuration in Fig.~\ref{Figure6}(e)--(h). Again, we first discuss the hot-carrier polarization in the emitter shown in Fig.~\ref{Figure6}(e). The profiles resemble the parallel case with a less developed decrease of the polarization in between the initial, optically excited maximum and the later increase due to pronounced back scattering and backflow of majority carriers from collector layer that has an antiparallel alignment. The transmission profile for the collector is exactly opposite to the previously discussed case, meaning that minority electrons easily enter and leave the collector, whereas majority carriers have a lower probability to do so. Note that we measure the spin polarization as positive for $\uparrow$ spins, which are the majority spins in the emitter. Fig.~\ref{Figure6}(f) shows, that in the collector, generation of minority electrons by scattering processes is more likely which explains the signal the strong negative polarization in the antiparallel configuration. Note, however, that there is a small positive polarization peak due to incoming $\uparrow$ electrons. Unlike in the parallel scenario, the signal recovers and shows a weak positive tail due to particles that propagated through the Au layer or the entire slab twice and reenter the collector. Thus we find for the averaged polarization in the collector for the antiparallel alignment a similar result to that of the parallel alignment, i.e., an enhanced hot-carrier polarization in the direction of the local magnetization. This result may be at first surprising, but will be explained below in connection with Fig.~\ref{Figure7}(c). 
	
	The spin-current in the Au layer, shown in Fig.~\ref{Figure6}(g) resembles the result of the parallel configuration but has a slightly higher magnitude, since fewer carriers are absorbed in the collector. This behavior already provides a hint that the main difference between the parallel and antiparallel configuration is the transport dynamics in the collector. This is corroborated by Fig.~\ref{Figure6}(h) which plots the spin-current evaluated at different positions of the collector. Here, we find a pronounced depth dependence, which is to be expected because primary $\uparrow$ carriers excited in the emitter scatter heavily in the $\downarrow$ collector, as can be seen from the inelastic mean-free paths in Fig.~\ref{Figure1} that differ by about a factor of 10 for hot minority and majority electrons in Fe. While the bipolar shape of the spin current close to the Au interface in the collector is similar to the result for the parallel case, it becomes strongly suppressed and switches its polarity with increasing penetration depth in the collector.
	
	We want to compare our result to Ref.~\onlinecite{Alekhin:2016wb}, where spin currents were measured for  a parallel and antiparallel configuration of a spin-valve structure, but Ref.~\onlinecite{Alekhin:2016wb} compared different magnetizations of the emitter, whereas we change the magnetization in the collector. They find that at short times the magnetization direction of the emitter determines the spin current mainly at the interfaces and at longer time scales ($t>500$\,fs) relaxation to quasi-equilibrium affects the magnetization. Our Fig.~\ref{Figure6}(d) and (h) show  that injecting carriers into a collector layer of the same or antiparallel magnetization does not change the induced spin current close to the interface and we believe that this in agreement with Ref.~\onlinecite{Alekhin:2016wb}. Our new result is that we can look into the layers, and the spin current in the parallel configuration decreases, but does not change sign, while in the antiparallel case, it strongly decreases and changes sign. 
	%Since Ref.~\onlinecite{Alekhin:2016wb} shows the spin current in the bulk of the collector layer, we would need to integrate the current over the depth of the collector and consequently, due to the discussion above, expect a spin current that does show the same polarization as the emitter layer. This is in agreement with the observation by Alekhin \textit{et al}.~\cite{Alekhin:2016wb}
	
	Figure~\ref{Figure7} shows the space- and time-resolved hot-carrier polarization and mean particle generation for the parallel and antiparallel configuration of a Fe(15nm)/Au(10nm)/Fe(15nm) spin-valve structure. For both alignments, the emitter (i.e., the left layer extending from 0 to 15\,nm) and the Au layer, extending from 15\,nm to 25\,nm show a similar behavior. The hot carrier polarization rises uniformly and is quickly injected into the Au where it dominates the dynamics for longer than 0.5ps. Both the polarization and the mean particle generation deviate strongly only in the collector (right layer extending from 25\,nm to 40\, nm). In the parallel case, the polarization is enhanced over the equilibrium polarization due to the favorable transmission of majority carriers at the Au-Fe interface. In addition these carriers act there as an additional pump exciting electrons from below the Fermi level. This latter process leads to a slow decrease of the polarization. Note that the polarization and generation profiles are largely spatially homogeneous in Fig.~\ref{Figure7}(a) and (b).
	
	In the antiparallel configuration, Fig.~\ref{Figure7}(c) and (d) the polarization and generation profiles in the collector show a pronounced depth dependence. While the hot-electron polarization in the first few nanometers is that of the injected primary carriers, it changes with increasing depth to the direction of the equilibrium magnetization of the collector. This behavior occurs because majority carriers originating from the emitter are less likely to enter the magnetic collector. When they do enter, they are strongly scattered and the dynamics in the layer is dominated by secondaries that carry the opposite polarization, i.e., a negative polarization in our notation. These electrons again can exit the layer easily but do not change the polarization in the Au layer appreciably due to their comparatively small density. This result shows the importance of depth-resolved measurements~\cite{Hofherr:2017dy}, since the carrier dynamics due to electron transport changes remarkably even on a nanometer scale. In Ref.~\onlinecite{Rudolf:1ky} for thin layered structures containing a 4\,nm collector, where an opposite behavior of the polarization dynamics for opposite alignments of emitter and collector was found. While the structure in Ref.~\onlinecite{Rudolf:1ky} is quite different from ours, in Fig.~\ref{Figure7}(c) we clearly also find such an opposite behavior, but only for a limited penetration depth. Only when we average the polarization over the 15\,nm collector we find an effective enhancement of the collector polarization both for parallel and antiparallel alignment of emitter and collector. 
	
	%%%%%%%%%%%%%%%
	% Conclusions %
	%%%%%%%%%%%%%%%
	\section{Conclusion}
	We presented a numerical study of hot-electron dynamics in ferromagnet-metal heterostructures after femtosecond optical excitation. Our approach is based on the Boltzmann transport equation with a scattering term that includes secondary carrier generation. The dynamical equations are solved by a particle-in-cell approach that yields classical equations of motion for superparticles sampling the electronic distribution function. We outlined an efficient numerical solution using the Strang operator splitting and showed that the model reproduces superdiffusive transport behavior correctly.
	
	For spin polarized transport in a Fe(15nm)/Au(x)-heterostructure, we find a fast rise of the hot-carrier induced polarization in the Au layer on the timescale that competes with ultrafast demagnetization processes. While the main effect on the hot-electron spin transport is due to optically excited primary carriers, secondary carrier generation plays a role in both materials and strongly influences the transport on longer timescales. We also analyzed a Fe/Au/Fe spin valve structure for parallel and antiparallel configurations. In this case, the interface transmission and the spin-dependent scattering in particular in the ferromagnetic collector, play a crucial role in the transport behavior. An important result for the antiparallel configuration is that injected minority electrons strongly scatter and the transport in the collector becomes strongly position dependent. In this configuration the simple picture of almost ballistic transport fails, as the scattered electrons effectively do not lead to a depolarization in the whole layer of the ferromagnet.
	As the results presented in this paper show, our model provides an intuitive and transparent picture of how ultrafast transport simulations can be done efficiently using \textit{ab initio} input and how these calculations help clarify  the dynamics of carriers measured in typical experimental set-ups.
	
	%%%%%%%%%%%%%%%%%%%
	% Acknowledgments %
	%%%%%%%%%%%%%%%%%%%
	\begin{acknowledgments}
		This work was supported by the German Science Foundation in the framework of the SFB/TRR 173 Spin+X (Project B03). DMN~acknowledges financial support from the Graduate School of Excellence MAINZ (Excellence Initiative DFG/GSC 266). DMN and HCS thank  R. Binder and the Optical Sciences Center at the University of Arizona for hospitality. 
	\end{acknowledgments}
	
	%%%%%%%%%%%%%%
	% References %
	%%%%%%%%%%%%%%
%	\bibliography{References}

\begin{thebibliography}{48}%
\makeatletter
\providecommand \@ifxundefined [1]{%
 \@ifx{#1\undefined}
}%
\providecommand \@ifnum [1]{%
 \ifnum #1\expandafter \@firstoftwo
 \else \expandafter \@secondoftwo
 \fi
}%
\providecommand \@ifx [1]{%
 \ifx #1\expandafter \@firstoftwo
 \else \expandafter \@secondoftwo
 \fi
}%
\providecommand \natexlab [1]{#1}%
\providecommand \enquote  [1]{``#1''}%
\providecommand \bibnamefont  [1]{#1}%
\providecommand \bibfnamefont [1]{#1}%
\providecommand \citenamefont [1]{#1}%
\providecommand \href@noop [0]{\@secondoftwo}%
\providecommand \href [0]{\begingroup \@sanitize@url \@href}%
\providecommand \@href[1]{\@@startlink{#1}\@@href}%
\providecommand \@@href[1]{\endgroup#1\@@endlink}%
\providecommand \@sanitize@url [0]{\catcode `\\12\catcode `\$12\catcode
  `\&12\catcode `\#12\catcode `\^12\catcode `\_12\catcode `\%12\relax}%
\providecommand \@@startlink[1]{}%
\providecommand \@@endlink[0]{}%
\providecommand \url  [0]{\begingroup\@sanitize@url \@url }%
\providecommand \@url [1]{\endgroup\@href {#1}{\urlprefix }}%
\providecommand \urlprefix  [0]{URL }%
\providecommand \Eprint [0]{\href }%
\providecommand \doibase [0]{http://dx.doi.org/}%
\providecommand \selectlanguage [0]{\@gobble}%
\providecommand \bibinfo  [0]{\@secondoftwo}%
\providecommand \bibfield  [0]{\@secondoftwo}%
\providecommand \translation [1]{[#1]}%
\providecommand \BibitemOpen [0]{}%
\providecommand \bibitemStop [0]{}%
\providecommand \bibitemNoStop [0]{.\EOS\space}%
\providecommand \EOS [0]{\spacefactor3000\relax}%
\providecommand \BibitemShut  [1]{\csname bibitem#1\endcsname}%
\let\auto@bib@innerbib\@empty
%</preamble>
\bibitem [{\citenamefont {Battiato}\ \emph {et~al.}(2010)\citenamefont
  {Battiato}, \citenamefont {Carva},\ and\ \citenamefont
  {Oppeneer}}]{Battiato:2010br}%
  \BibitemOpen
  \bibfield  {author} {\bibinfo {author} {\bibfnamefont {M.}~\bibnamefont
  {Battiato}}, \bibinfo {author} {\bibfnamefont {K.}~\bibnamefont {Carva}}, \
  and\ \bibinfo {author} {\bibfnamefont {P.~M.}\ \bibnamefont {Oppeneer}},\
  }\href@noop {} {\bibfield  {journal} {\bibinfo  {journal} {Physical Review
  Letters}\ }\textbf {\bibinfo {volume} {105}},\ \bibinfo {pages} {027203}
  (\bibinfo {year} {2010})}\BibitemShut {NoStop}%
\bibitem [{\citenamefont {Rudolf}\ \emph {et~al.}(1)\citenamefont {Rudolf},
  \citenamefont {La-O-Vorakiat}, \citenamefont {Battiato}, \citenamefont
  {Adam}, \citenamefont {Shaw}, \citenamefont {Turgut}, \citenamefont
  {Maldonado}, \citenamefont {Mathias}, \citenamefont {Grychtol}, \citenamefont
  {Nembach}, \citenamefont {Silva}, \citenamefont {Aeschlimann}, \citenamefont
  {Kapteyn}, \citenamefont {Murnane}, \citenamefont {Schneider},\ and\
  \citenamefont {Oppeneer}}]{Rudolf:1ky}%
  \BibitemOpen
  \bibfield  {author} {\bibinfo {author} {\bibfnamefont {D.}~\bibnamefont
  {Rudolf}}, \bibinfo {author} {\bibfnamefont {C.}~\bibnamefont
  {La-O-Vorakiat}}, \bibinfo {author} {\bibfnamefont {M.}~\bibnamefont
  {Battiato}}, \bibinfo {author} {\bibfnamefont {R.}~\bibnamefont {Adam}},
  \bibinfo {author} {\bibfnamefont {J.~M.}\ \bibnamefont {Shaw}}, \bibinfo
  {author} {\bibfnamefont {E.}~\bibnamefont {Turgut}}, \bibinfo {author}
  {\bibfnamefont {P.}~\bibnamefont {Maldonado}}, \bibinfo {author}
  {\bibfnamefont {S.}~\bibnamefont {Mathias}}, \bibinfo {author} {\bibfnamefont
  {P.}~\bibnamefont {Grychtol}}, \bibinfo {author} {\bibfnamefont {H.~T.}\
  \bibnamefont {Nembach}}, \bibinfo {author} {\bibfnamefont {T.~J.}\
  \bibnamefont {Silva}}, \bibinfo {author} {\bibfnamefont {M.}~\bibnamefont
  {Aeschlimann}}, \bibinfo {author} {\bibfnamefont {H.~C.}\ \bibnamefont
  {Kapteyn}}, \bibinfo {author} {\bibfnamefont {M.~M.}\ \bibnamefont
  {Murnane}}, \bibinfo {author} {\bibfnamefont {C.~M.}\ \bibnamefont
  {Schneider}}, \ and\ \bibinfo {author} {\bibfnamefont {P.~M.}\ \bibnamefont
  {Oppeneer}},\ }\href@noop {} {\bibfield  {journal} {\bibinfo  {journal}
  {Nature Communications}\ }\textbf {\bibinfo {volume} {3}},\ \bibinfo {pages}
  {1037} (\bibinfo {year} {1})}\BibitemShut {NoStop}%
\bibitem [{\citenamefont {Bergeard}\ \emph {et~al.}(2016)\citenamefont
  {Bergeard}, \citenamefont {Hehn}, \citenamefont {Mangin}, \citenamefont
  {Lengaigne}, \citenamefont {Montaigne}, \citenamefont {Lalieu}, \citenamefont
  {Koopmans},\ and\ \citenamefont {Malinowski}}]{Bergeard:2016jka}%
  \BibitemOpen
  \bibfield  {author} {\bibinfo {author} {\bibfnamefont {N.}~\bibnamefont
  {Bergeard}}, \bibinfo {author} {\bibfnamefont {M.}~\bibnamefont {Hehn}},
  \bibinfo {author} {\bibfnamefont {S.}~\bibnamefont {Mangin}}, \bibinfo
  {author} {\bibfnamefont {G.}~\bibnamefont {Lengaigne}}, \bibinfo {author}
  {\bibfnamefont {F.}~\bibnamefont {Montaigne}}, \bibinfo {author}
  {\bibfnamefont {M.~L.~M.}\ \bibnamefont {Lalieu}}, \bibinfo {author}
  {\bibfnamefont {B.}~\bibnamefont {Koopmans}}, \ and\ \bibinfo {author}
  {\bibfnamefont {G.}~\bibnamefont {Malinowski}},\ }\href@noop {} {\bibfield
  {journal} {\bibinfo  {journal} {Physical Review Letters}\ }\textbf {\bibinfo
  {volume} {117}},\ \bibinfo {pages} {147203} (\bibinfo {year}
  {2016})}\BibitemShut {NoStop}%
\bibitem [{\citenamefont {Salvatella}\ \emph {et~al.}(2016)\citenamefont
  {Salvatella}, \citenamefont {Gort}, \citenamefont {B{\"u}hlmann},
  \citenamefont {D{\"a}ster}, \citenamefont {Vaterlaus},\ and\ \citenamefont
  {Acremann}}]{Salvatella:2016gz}%
  \BibitemOpen
  \bibfield  {author} {\bibinfo {author} {\bibfnamefont {G.}~\bibnamefont
  {Salvatella}}, \bibinfo {author} {\bibfnamefont {R.}~\bibnamefont {Gort}},
  \bibinfo {author} {\bibfnamefont {K.}~\bibnamefont {B{\"u}hlmann}}, \bibinfo
  {author} {\bibfnamefont {S.}~\bibnamefont {D{\"a}ster}}, \bibinfo {author}
  {\bibfnamefont {A.}~\bibnamefont {Vaterlaus}}, \ and\ \bibinfo {author}
  {\bibfnamefont {Y.}~\bibnamefont {Acremann}},\ }\href@noop {} {\bibfield
  {journal} {\bibinfo  {journal} {Structural Dynamics}\ }\textbf {\bibinfo
  {volume} {3}},\ \bibinfo {pages} {055101} (\bibinfo {year}
  {2016})}\BibitemShut {NoStop}%
\bibitem [{\citenamefont {Turgut}\ \emph {et~al.}(2016)\citenamefont {Turgut},
  \citenamefont {Zusin}, \citenamefont {Legut}, \citenamefont {Carva},
  \citenamefont {Knut}, \citenamefont {Shaw}, \citenamefont {Chen},
  \citenamefont {Tao}, \citenamefont {Nembach}, \citenamefont {Silva},
  \citenamefont {Mathias}, \citenamefont {Aeschlimann}, \citenamefont
  {Oppeneer}, \citenamefont {Kapteyn}, \citenamefont {Murnane},\ and\
  \citenamefont {Grychtol}}]{Turgut:2016dx}%
  \BibitemOpen
  \bibfield  {author} {\bibinfo {author} {\bibfnamefont {E.}~\bibnamefont
  {Turgut}}, \bibinfo {author} {\bibfnamefont {D.}~\bibnamefont {Zusin}},
  \bibinfo {author} {\bibfnamefont {D.}~\bibnamefont {Legut}}, \bibinfo
  {author} {\bibfnamefont {K.}~\bibnamefont {Carva}}, \bibinfo {author}
  {\bibfnamefont {R.}~\bibnamefont {Knut}}, \bibinfo {author} {\bibfnamefont
  {J.~M.}\ \bibnamefont {Shaw}}, \bibinfo {author} {\bibfnamefont
  {C.}~\bibnamefont {Chen}}, \bibinfo {author} {\bibfnamefont {Z.}~\bibnamefont
  {Tao}}, \bibinfo {author} {\bibfnamefont {H.~T.}\ \bibnamefont {Nembach}},
  \bibinfo {author} {\bibfnamefont {T.~J.}\ \bibnamefont {Silva}}, \bibinfo
  {author} {\bibfnamefont {S.}~\bibnamefont {Mathias}}, \bibinfo {author}
  {\bibfnamefont {M.}~\bibnamefont {Aeschlimann}}, \bibinfo {author}
  {\bibfnamefont {P.~M.}\ \bibnamefont {Oppeneer}}, \bibinfo {author}
  {\bibfnamefont {H.~C.}\ \bibnamefont {Kapteyn}}, \bibinfo {author}
  {\bibfnamefont {M.~M.}\ \bibnamefont {Murnane}}, \ and\ \bibinfo {author}
  {\bibfnamefont {P.}~\bibnamefont {Grychtol}},\ }\href@noop {} {\bibfield
  {journal} {\bibinfo  {journal} {Physical Review B}\ }\textbf {\bibinfo
  {volume} {94}},\ \bibinfo {pages} {220408} (\bibinfo {year}
  {2016})}\BibitemShut {NoStop}%
\bibitem [{\citenamefont {Bigot}\ \emph {et~al.}(2009)\citenamefont {Bigot},
  \citenamefont {Vomir},\ and\ \citenamefont
  {Beaurepaire}}]{bigot2009coherent}%
  \BibitemOpen
  \bibfield  {author} {\bibinfo {author} {\bibfnamefont {J.-Y.}\ \bibnamefont
  {Bigot}}, \bibinfo {author} {\bibfnamefont {M.}~\bibnamefont {Vomir}}, \ and\
  \bibinfo {author} {\bibfnamefont {E.}~\bibnamefont {Beaurepaire}},\
  }\href@noop {} {\bibfield  {journal} {\bibinfo  {journal} {Nature Physics}\
  }\textbf {\bibinfo {volume} {5}},\ \bibinfo {pages} {515} (\bibinfo {year}
  {2009})}\BibitemShut {NoStop}%
\bibitem [{\citenamefont {Melnikov}\ \emph
  {et~al.}(2011{\natexlab{a}})\citenamefont {Melnikov}, \citenamefont
  {Razdolski}, \citenamefont {Wehling}, \citenamefont {Papaioannou},
  \citenamefont {Roddatis}, \citenamefont {Fumagalli}, \citenamefont
  {Aktsipetrov}, \citenamefont {Lichtenstein},\ and\ \citenamefont
  {Bovensiepen}}]{Melnikov:2011ep}%
  \BibitemOpen
  \bibfield  {author} {\bibinfo {author} {\bibfnamefont {A.}~\bibnamefont
  {Melnikov}}, \bibinfo {author} {\bibfnamefont {I.}~\bibnamefont {Razdolski}},
  \bibinfo {author} {\bibfnamefont {T.~O.}\ \bibnamefont {Wehling}}, \bibinfo
  {author} {\bibfnamefont {E.~T.}\ \bibnamefont {Papaioannou}}, \bibinfo
  {author} {\bibfnamefont {V.}~\bibnamefont {Roddatis}}, \bibinfo {author}
  {\bibfnamefont {P.}~\bibnamefont {Fumagalli}}, \bibinfo {author}
  {\bibfnamefont {O.}~\bibnamefont {Aktsipetrov}}, \bibinfo {author}
  {\bibfnamefont {A.~I.}\ \bibnamefont {Lichtenstein}}, \ and\ \bibinfo
  {author} {\bibfnamefont {U.}~\bibnamefont {Bovensiepen}},\ }\href@noop {}
  {\bibfield  {journal} {\bibinfo  {journal} {Physical Review Letters}\
  }\textbf {\bibinfo {volume} {107}},\ \bibinfo {pages} {076601} (\bibinfo
  {year} {2011}{\natexlab{a}})}\BibitemShut {NoStop}%
\bibitem [{\citenamefont {Schellekens}\ \emph {et~al.}(2014)\citenamefont
  {Schellekens}, \citenamefont {de~Vries}, \citenamefont {Lucassen},\ and\
  \citenamefont {Koopmans}}]{Schellekens2014}%
  \BibitemOpen
  \bibfield  {author} {\bibinfo {author} {\bibfnamefont {A.~J.}\ \bibnamefont
  {Schellekens}}, \bibinfo {author} {\bibfnamefont {N.}~\bibnamefont
  {de~Vries}}, \bibinfo {author} {\bibfnamefont {J.}~\bibnamefont {Lucassen}},
  \ and\ \bibinfo {author} {\bibfnamefont {B.}~\bibnamefont {Koopmans}},\
  }\href {\doibase 10.1103/PhysRevB.90.104429} {\bibfield  {journal} {\bibinfo
  {journal} {Phys. Rev. B}\ }\textbf {\bibinfo {volume} {90}},\ \bibinfo
  {pages} {104429} (\bibinfo {year} {2014})}\BibitemShut {NoStop}%
\bibitem [{\citenamefont {Hofherr}\ \emph
  {et~al.}(2017{\natexlab{a}})\citenamefont {Hofherr}, \citenamefont
  {Maldonado}, \citenamefont {Schmitt}, \citenamefont {Berritta}, \citenamefont
  {Bierbrauer}, \citenamefont {Sadashivaiah}, \citenamefont {Schellekens},
  \citenamefont {Koopmans}, \citenamefont {Steil}, \citenamefont {Cinchetti},
  \citenamefont {Stadtm{\"u}ller}, \citenamefont {Oppeneer}, \citenamefont
  {Mathias},\ and\ \citenamefont {Aeschlimann}}]{Hofherr:2017dya}%
  \BibitemOpen
  \bibfield  {author} {\bibinfo {author} {\bibfnamefont {M.}~\bibnamefont
  {Hofherr}}, \bibinfo {author} {\bibfnamefont {P.}~\bibnamefont {Maldonado}},
  \bibinfo {author} {\bibfnamefont {O.}~\bibnamefont {Schmitt}}, \bibinfo
  {author} {\bibfnamefont {M.}~\bibnamefont {Berritta}}, \bibinfo {author}
  {\bibfnamefont {U.}~\bibnamefont {Bierbrauer}}, \bibinfo {author}
  {\bibfnamefont {S.}~\bibnamefont {Sadashivaiah}}, \bibinfo {author}
  {\bibfnamefont {A.~J.}\ \bibnamefont {Schellekens}}, \bibinfo {author}
  {\bibfnamefont {B.}~\bibnamefont {Koopmans}}, \bibinfo {author}
  {\bibfnamefont {D.}~\bibnamefont {Steil}}, \bibinfo {author} {\bibfnamefont
  {M.}~\bibnamefont {Cinchetti}}, \bibinfo {author} {\bibfnamefont
  {B.}~\bibnamefont {Stadtm{\"u}ller}}, \bibinfo {author} {\bibfnamefont
  {P.~M.}\ \bibnamefont {Oppeneer}}, \bibinfo {author} {\bibfnamefont
  {S.}~\bibnamefont {Mathias}}, \ and\ \bibinfo {author} {\bibfnamefont
  {M.}~\bibnamefont {Aeschlimann}},\ }\href@noop {} {\bibfield  {journal}
  {\bibinfo  {journal} {Physical Review B}\ }\textbf {\bibinfo {volume} {96}},\
  \bibinfo {pages} {100403} (\bibinfo {year} {2017}{\natexlab{a}})}\BibitemShut
  {NoStop}%
\bibitem [{\citenamefont {Valet}\ and\ \citenamefont
  {Fert}(1993{\natexlab{a}})}]{Valet1993}%
  \BibitemOpen
  \bibfield  {author} {\bibinfo {author} {\bibfnamefont {T.}~\bibnamefont
  {Valet}}\ and\ \bibinfo {author} {\bibfnamefont {A.}~\bibnamefont {Fert}},\
  }\href@noop {} {\bibfield  {journal} {\bibinfo  {journal} {Phys. Rev. B
  Condens. Matter}\ }\textbf {\bibinfo {volume} {48}},\ \bibinfo {pages} {7099}
  (\bibinfo {year} {1993}{\natexlab{a}})}\BibitemShut {NoStop}%
\bibitem [{\citenamefont {Zhang}\ and\ \citenamefont {Levy}(2002)}]{Zhang2002}%
  \BibitemOpen
  \bibfield  {author} {\bibinfo {author} {\bibfnamefont {S.}~\bibnamefont
  {Zhang}}\ and\ \bibinfo {author} {\bibfnamefont {P.~M.}\ \bibnamefont
  {Levy}},\ }\href {\doibase 10.1103/PhysRevB.65.052409} {\bibfield  {journal}
  {\bibinfo  {journal} {Phys. Rev. B}\ }\textbf {\bibinfo {volume} {65}},\
  \bibinfo {pages} {052409} (\bibinfo {year} {2002})}\BibitemShut {NoStop}%
\bibitem [{\citenamefont {Xiao}\ \emph {et~al.}(2007)\citenamefont {Xiao},
  \citenamefont {Zangwill},\ and\ \citenamefont {Stiles}}]{Xiao2007}%
  \BibitemOpen
  \bibfield  {author} {\bibinfo {author} {\bibfnamefont {J.}~\bibnamefont
  {Xiao}}, \bibinfo {author} {\bibfnamefont {A.}~\bibnamefont {Zangwill}}, \
  and\ \bibinfo {author} {\bibfnamefont {M.~D.}\ \bibnamefont {Stiles}},\
  }\href@noop {} {\bibfield  {journal} {\bibinfo  {journal} {Eur. Phys. J. B}\
  }\textbf {\bibinfo {volume} {59}},\ \bibinfo {pages} {415} (\bibinfo {year}
  {2007})}\BibitemShut {NoStop}%
\bibitem [{\citenamefont {Butler}(1985)}]{Butler1985}%
  \BibitemOpen
  \bibfield  {author} {\bibinfo {author} {\bibfnamefont {W.~H.}\ \bibnamefont
  {Butler}},\ }\href {\doibase 10.1103/PhysRevB.31.3260} {\bibfield  {journal}
  {\bibinfo  {journal} {Phys. Rev. B}\ }\textbf {\bibinfo {volume} {31}},\
  \bibinfo {pages} {3260} (\bibinfo {year} {1985})}\BibitemShut {NoStop}%
\bibitem [{\citenamefont {Zhang}\ \emph {et~al.}(1992)\citenamefont {Zhang},
  \citenamefont {Levy},\ and\ \citenamefont {Fert}}]{Zhang1992}%
  \BibitemOpen
  \bibfield  {author} {\bibinfo {author} {\bibfnamefont {S.}~\bibnamefont
  {Zhang}}, \bibinfo {author} {\bibfnamefont {P.~M.}\ \bibnamefont {Levy}}, \
  and\ \bibinfo {author} {\bibfnamefont {A.}~\bibnamefont {Fert}},\ }\href
  {\doibase 10.1103/PhysRevB.45.8689} {\bibfield  {journal} {\bibinfo
  {journal} {Phys. Rev. B}\ }\textbf {\bibinfo {volume} {45}},\ \bibinfo
  {pages} {8689} (\bibinfo {year} {1992})}\BibitemShut {NoStop}%
\bibitem [{\citenamefont {Kaltenborn}\ \emph {et~al.}(2012)\citenamefont
  {Kaltenborn}, \citenamefont {Zhu},\ and\ \citenamefont
  {Schneider}}]{Kaltenborn:2014hg}%
  \BibitemOpen
  \bibfield  {author} {\bibinfo {author} {\bibfnamefont {S.}~\bibnamefont
  {Kaltenborn}}, \bibinfo {author} {\bibfnamefont {Y.-H.}\ \bibnamefont {Zhu}},
  \ and\ \bibinfo {author} {\bibfnamefont {H.~C.}\ \bibnamefont {Schneider}},\
  }\href {\doibase 10.1103/PhysRevB.85.235101} {\bibfield  {journal} {\bibinfo
  {journal} {Phys. Rev. B}\ }\textbf {\bibinfo {volume} {85}},\ \bibinfo
  {pages} {235101} (\bibinfo {year} {2012})}\BibitemShut {NoStop}%
\bibitem [{\citenamefont {Huthmacher}\ \emph {et~al.}(2016)\citenamefont
  {Huthmacher}, \citenamefont {Molberg}, \citenamefont {Rethfeld},\ and\
  \citenamefont {Gulley}}]{Huthmacher:2016dd}%
  \BibitemOpen
  \bibfield  {author} {\bibinfo {author} {\bibfnamefont {K.}~\bibnamefont
  {Huthmacher}}, \bibinfo {author} {\bibfnamefont {A.~K.}\ \bibnamefont
  {Molberg}}, \bibinfo {author} {\bibfnamefont {B.}~\bibnamefont {Rethfeld}}, \
  and\ \bibinfo {author} {\bibfnamefont {J.~R.}\ \bibnamefont {Gulley}},\
  }\href@noop {} {\bibfield  {journal} {\bibinfo  {journal} {Journal of
  Computational Physics}\ }\textbf {\bibinfo {volume} {322}},\ \bibinfo {pages}
  {535} (\bibinfo {year} {2016})}\BibitemShut {NoStop}%
\bibitem [{\citenamefont {Wieczorek}\ \emph {et~al.}(2015)\citenamefont
  {Wieczorek}, \citenamefont {Eschenlohr}, \citenamefont {Weidtmann},
  \citenamefont {R{\"o}sner}, \citenamefont {Bergeard}, \citenamefont
  {Tarasevitch}, \citenamefont {Wehling},\ and\ \citenamefont
  {Bovensiepen}}]{Wieczorek:2015fk}%
  \BibitemOpen
  \bibfield  {author} {\bibinfo {author} {\bibfnamefont {J.}~\bibnamefont
  {Wieczorek}}, \bibinfo {author} {\bibfnamefont {A.}~\bibnamefont
  {Eschenlohr}}, \bibinfo {author} {\bibfnamefont {B.}~\bibnamefont
  {Weidtmann}}, \bibinfo {author} {\bibfnamefont {M.}~\bibnamefont
  {R{\"o}sner}}, \bibinfo {author} {\bibfnamefont {N.}~\bibnamefont
  {Bergeard}}, \bibinfo {author} {\bibfnamefont {A.}~\bibnamefont
  {Tarasevitch}}, \bibinfo {author} {\bibfnamefont {T.~O.}\ \bibnamefont
  {Wehling}}, \ and\ \bibinfo {author} {\bibfnamefont {U.}~\bibnamefont
  {Bovensiepen}},\ }\href@noop {} {\bibfield  {journal} {\bibinfo  {journal}
  {Physical Review B}\ }\textbf {\bibinfo {volume} {92}},\ \bibinfo {pages}
  {174410} (\bibinfo {year} {2015})}\BibitemShut {NoStop}%
\bibitem [{\citenamefont {Kampfrath}\ \emph {et~al.}(2013)\citenamefont
  {Kampfrath}, \citenamefont {Battiato}, \citenamefont {Maldonado},
  \citenamefont {Eilers}, \citenamefont {N{\"o}tzold}, \citenamefont
  {M{\"a}hrlein}, \citenamefont {Zbarsky}, \citenamefont {Freimuth},
  \citenamefont {Mokrousov}, \citenamefont {Bl{\"u}gel}, \citenamefont {Wolf},
  \citenamefont {Radu}, \citenamefont {Oppeneer},\ and\ \citenamefont
  {M{\"u}nzenberg}}]{Kampfrath:2013kw}%
  \BibitemOpen
  \bibfield  {author} {\bibinfo {author} {\bibfnamefont {T.}~\bibnamefont
  {Kampfrath}}, \bibinfo {author} {\bibfnamefont {M.}~\bibnamefont {Battiato}},
  \bibinfo {author} {\bibfnamefont {P.}~\bibnamefont {Maldonado}}, \bibinfo
  {author} {\bibfnamefont {G.}~\bibnamefont {Eilers}}, \bibinfo {author}
  {\bibfnamefont {J.}~\bibnamefont {N{\"o}tzold}}, \bibinfo {author}
  {\bibfnamefont {S.}~\bibnamefont {M{\"a}hrlein}}, \bibinfo {author}
  {\bibfnamefont {V.}~\bibnamefont {Zbarsky}}, \bibinfo {author} {\bibfnamefont
  {F.}~\bibnamefont {Freimuth}}, \bibinfo {author} {\bibfnamefont
  {Y.}~\bibnamefont {Mokrousov}}, \bibinfo {author} {\bibfnamefont
  {S.}~\bibnamefont {Bl{\"u}gel}}, \bibinfo {author} {\bibfnamefont
  {M.}~\bibnamefont {Wolf}}, \bibinfo {author} {\bibfnamefont {I.}~\bibnamefont
  {Radu}}, \bibinfo {author} {\bibfnamefont {P.~M.}\ \bibnamefont {Oppeneer}},
  \ and\ \bibinfo {author} {\bibfnamefont {M.}~\bibnamefont {M{\"u}nzenberg}},\
  }\href@noop {} {\bibfield  {journal} {\bibinfo  {journal} {Nature
  Nanotechnology}\ }\textbf {\bibinfo {volume} {8}},\ \bibinfo {pages} {256}
  (\bibinfo {year} {2013})}\BibitemShut {NoStop}%
\bibitem [{\citenamefont {Eschenlohr}\ \emph {et~al.}(2013)\citenamefont
  {Eschenlohr}, \citenamefont {Battiato}, \citenamefont {Maldonado},
  \citenamefont {Pontius}, \citenamefont {Kachel}, \citenamefont {Holldack},
  \citenamefont {Mitzner}, \citenamefont {F{\"o}hlisch}, \citenamefont
  {Oppeneer},\ and\ \citenamefont {Stamm}}]{Eschenlohr:2013id}%
  \BibitemOpen
  \bibfield  {author} {\bibinfo {author} {\bibfnamefont {A.}~\bibnamefont
  {Eschenlohr}}, \bibinfo {author} {\bibfnamefont {M.}~\bibnamefont
  {Battiato}}, \bibinfo {author} {\bibfnamefont {P.}~\bibnamefont {Maldonado}},
  \bibinfo {author} {\bibfnamefont {N.}~\bibnamefont {Pontius}}, \bibinfo
  {author} {\bibfnamefont {T.}~\bibnamefont {Kachel}}, \bibinfo {author}
  {\bibfnamefont {K.}~\bibnamefont {Holldack}}, \bibinfo {author}
  {\bibfnamefont {R.}~\bibnamefont {Mitzner}}, \bibinfo {author} {\bibfnamefont
  {A.}~\bibnamefont {F{\"o}hlisch}}, \bibinfo {author} {\bibfnamefont {P.~M.}\
  \bibnamefont {Oppeneer}}, \ and\ \bibinfo {author} {\bibfnamefont
  {C.}~\bibnamefont {Stamm}},\ }\href@noop {} {\bibfield  {journal} {\bibinfo
  {journal} {Nature materials}\ }\textbf {\bibinfo {volume} {12}},\ \bibinfo
  {pages} {332} (\bibinfo {year} {2013})}\BibitemShut {NoStop}%
\bibitem [{\citenamefont {Battiato}\ \emph {et~al.}(2014)\citenamefont
  {Battiato}, \citenamefont {Maldonado},\ and\ \citenamefont
  {Oppeneer}}]{Battiato:2014kz}%
  \BibitemOpen
  \bibfield  {author} {\bibinfo {author} {\bibfnamefont {M.}~\bibnamefont
  {Battiato}}, \bibinfo {author} {\bibfnamefont {P.}~\bibnamefont {Maldonado}},
  \ and\ \bibinfo {author} {\bibfnamefont {P.~M.}\ \bibnamefont {Oppeneer}},\
  }\href@noop {} {\bibfield  {journal} {\bibinfo  {journal} {Journal of Applied
  Physics}\ }\textbf {\bibinfo {volume} {115}},\ \bibinfo {pages} {172611}
  (\bibinfo {year} {2014})}\BibitemShut {NoStop}%
\bibitem [{\citenamefont {Silin}(1957)}]{Silin:1957ko}%
  \BibitemOpen
  \bibfield  {author} {\bibinfo {author} {\bibfnamefont {V.~P.}\ \bibnamefont
  {Silin}},\ }\href@noop {} {\bibfield  {journal} {\bibinfo  {journal} {J.
  Exptl. Theoret. Phys. (U.S.S.R.)}\ }\textbf {\bibinfo {volume} {33}},\
  \bibinfo {pages} {495} (\bibinfo {year} {1957})}\BibitemShut {NoStop}%
\bibitem [{\citenamefont {Bonitz}(2016)}]{Bonitz:2015up}%
  \BibitemOpen
  \bibfield  {author} {\bibinfo {author} {\bibfnamefont {M.}~\bibnamefont
  {Bonitz}},\ }\href@noop {} {\emph {\bibinfo {title} {Quantum kinetic
  theory}}}\ (\bibinfo  {publisher} {Springer},\ \bibinfo {year}
  {2016})\BibitemShut {NoStop}%
\bibitem [{\citenamefont {Qi}\ and\ \citenamefont {Zhang}(2003)}]{Qi:2003ks}%
  \BibitemOpen
  \bibfield  {author} {\bibinfo {author} {\bibfnamefont {Y.}~\bibnamefont
  {Qi}}\ and\ \bibinfo {author} {\bibfnamefont {S.}~\bibnamefont {Zhang}},\
  }\href@noop {} {\bibfield  {journal} {\bibinfo  {journal} {Physical Review
  B}\ }\textbf {\bibinfo {volume} {67}},\ \bibinfo {pages} {052407} (\bibinfo
  {year} {2003})}\BibitemShut {NoStop}%
\bibitem [{\citenamefont {Valet}\ and\ \citenamefont
  {Fert}(1993{\natexlab{b}})}]{Valet:1993es}%
  \BibitemOpen
  \bibfield  {author} {\bibinfo {author} {\bibfnamefont {T.}~\bibnamefont
  {Valet}}\ and\ \bibinfo {author} {\bibfnamefont {A.}~\bibnamefont {Fert}},\
  }\href@noop {} {\bibfield  {journal} {\bibinfo  {journal} {Physical Review
  B}\ }\textbf {\bibinfo {volume} {48}},\ \bibinfo {pages} {7099} (\bibinfo
  {year} {1993}{\natexlab{b}})}\BibitemShut {NoStop}%
\bibitem [{\citenamefont {Zhu}\ \emph {et~al.}(2009)\citenamefont {Zhu},
  \citenamefont {Hillebrands},\ and\ \citenamefont {Schneider}}]{Zhu:2009dm}%
  \BibitemOpen
  \bibfield  {author} {\bibinfo {author} {\bibfnamefont {Y.-H.}\ \bibnamefont
  {Zhu}}, \bibinfo {author} {\bibfnamefont {B.}~\bibnamefont {Hillebrands}}, \
  and\ \bibinfo {author} {\bibfnamefont {H.~C.}\ \bibnamefont {Schneider}},\
  }\href@noop {} {\bibfield  {journal} {\bibinfo  {journal} {Physical Review
  B}\ }\textbf {\bibinfo {volume} {79}},\ \bibinfo {pages} {214412} (\bibinfo
  {year} {2009})}\BibitemShut {NoStop}%
\bibitem [{\citenamefont {Manfredi}\ and\ \citenamefont
  {Hervieux}(2005)}]{Manfredi:2005ba}%
  \BibitemOpen
  \bibfield  {author} {\bibinfo {author} {\bibfnamefont {G.}~\bibnamefont
  {Manfredi}}\ and\ \bibinfo {author} {\bibfnamefont {P.~A.}\ \bibnamefont
  {Hervieux}},\ }\href@noop {} {\bibfield  {journal} {\bibinfo  {journal}
  {Physical Review B}\ }\textbf {\bibinfo {volume} {72}},\ \bibinfo {pages}
  {155421} (\bibinfo {year} {2005})}\BibitemShut {NoStop}%
\bibitem [{\citenamefont {Hurst}\ \emph {et~al.}(2018)\citenamefont {Hurst},
  \citenamefont {Hervieux},\ and\ \citenamefont
  {Manfredi}}]{PhysRevB.97.014424}%
  \BibitemOpen
  \bibfield  {author} {\bibinfo {author} {\bibfnamefont {J.}~\bibnamefont
  {Hurst}}, \bibinfo {author} {\bibfnamefont {P.-A.}\ \bibnamefont {Hervieux}},
  \ and\ \bibinfo {author} {\bibfnamefont {G.}~\bibnamefont {Manfredi}},\
  }\href {\doibase 10.1103/PhysRevB.97.014424} {\bibfield  {journal} {\bibinfo
  {journal} {Phys. Rev. B}\ }\textbf {\bibinfo {volume} {97}},\ \bibinfo
  {pages} {014424} (\bibinfo {year} {2018})}\BibitemShut {NoStop}%
\bibitem [{\citenamefont {Wolff}(1954)}]{Wolff:1954ds}%
  \BibitemOpen
  \bibfield  {author} {\bibinfo {author} {\bibfnamefont {P.~A.}\ \bibnamefont
  {Wolff}},\ }\href@noop {} {\bibfield  {journal} {\bibinfo  {journal}
  {Physical Review}\ }\textbf {\bibinfo {volume} {95}},\ \bibinfo {pages} {56}
  (\bibinfo {year} {1954})}\BibitemShut {NoStop}%
\bibitem [{\citenamefont {Penn}\ \emph
  {et~al.}(1985{\natexlab{a}})\citenamefont {Penn}, \citenamefont {Apell},\
  and\ \citenamefont {Girvin}}]{Penn:1985us}%
  \BibitemOpen
  \bibfield  {author} {\bibinfo {author} {\bibfnamefont {D.~R.}\ \bibnamefont
  {Penn}}, \bibinfo {author} {\bibfnamefont {S.~P.}\ \bibnamefont {Apell}}, \
  and\ \bibinfo {author} {\bibfnamefont {S.~M.}\ \bibnamefont {Girvin}},\
  }\href {\doibase 10.1103/PhysRevLett.55.518} {\bibfield  {journal} {\bibinfo
  {journal} {Phys. Rev. Lett.}\ }\textbf {\bibinfo {volume} {55}},\ \bibinfo
  {pages} {518} (\bibinfo {year} {1985}{\natexlab{a}})}\BibitemShut {NoStop}%
\bibitem [{\citenamefont {Penn}\ \emph
  {et~al.}(1985{\natexlab{b}})\citenamefont {Penn}, \citenamefont {Apell},\
  and\ \citenamefont {Girvin}}]{Penn:1985wt}%
  \BibitemOpen
  \bibfield  {author} {\bibinfo {author} {\bibfnamefont {D.~R.}\ \bibnamefont
  {Penn}}, \bibinfo {author} {\bibfnamefont {S.~P.}\ \bibnamefont {Apell}}, \
  and\ \bibinfo {author} {\bibfnamefont {S.~M.}\ \bibnamefont {Girvin}},\
  }\href {\doibase 10.1103/PhysRevB.32.7753} {\bibfield  {journal} {\bibinfo
  {journal} {Phys. Rev. B}\ }\textbf {\bibinfo {volume} {32}},\ \bibinfo
  {pages} {7753} (\bibinfo {year} {1985}{\natexlab{b}})}\BibitemShut {NoStop}%
\bibitem [{\citenamefont {Zhukov}\ \emph {et~al.}(2006)\citenamefont {Zhukov},
  \citenamefont {Chulkov},\ and\ \citenamefont {Echenique}}]{Zhukov:2006ky}%
  \BibitemOpen
  \bibfield  {author} {\bibinfo {author} {\bibfnamefont {V.~P.}\ \bibnamefont
  {Zhukov}}, \bibinfo {author} {\bibfnamefont {E.~V.}\ \bibnamefont {Chulkov}},
  \ and\ \bibinfo {author} {\bibfnamefont {P.~M.}\ \bibnamefont {Echenique}},\
  }\href@noop {} {\bibfield  {journal} {\bibinfo  {journal} {Physical Review
  B}\ }\textbf {\bibinfo {volume} {73}},\ \bibinfo {pages} {465} (\bibinfo
  {year} {2006})}\BibitemShut {NoStop}%
\bibitem [{\citenamefont {Kaltenborn}\ and\ \citenamefont
  {Schneider}(2014)}]{Kaltenborn:2014du}%
  \BibitemOpen
  \bibfield  {author} {\bibinfo {author} {\bibfnamefont {S.}~\bibnamefont
  {Kaltenborn}}\ and\ \bibinfo {author} {\bibfnamefont {H.~C.}\ \bibnamefont
  {Schneider}},\ }\href@noop {} {\bibfield  {journal} {\bibinfo  {journal}
  {Physical Review B}\ }\textbf {\bibinfo {volume} {90}},\ \bibinfo {pages}
  {201104} (\bibinfo {year} {2014})}\BibitemShut {NoStop}%
\bibitem [{\citenamefont {Alekhin}\ \emph {et~al.}(2017)\citenamefont
  {Alekhin}, \citenamefont {Razdolski}, \citenamefont {Ilin}, \citenamefont
  {Meyburg}, \citenamefont {Diesing}, \citenamefont {Roddatis}, \citenamefont
  {Rungger}, \citenamefont {Stamenova}, \citenamefont {Sanvito}, \citenamefont
  {Bovensiepen},\ and\ \citenamefont {Melnikov}}]{Alekhin:2016wb}%
  \BibitemOpen
  \bibfield  {author} {\bibinfo {author} {\bibfnamefont {A.}~\bibnamefont
  {Alekhin}}, \bibinfo {author} {\bibfnamefont {I.}~\bibnamefont {Razdolski}},
  \bibinfo {author} {\bibfnamefont {N.}~\bibnamefont {Ilin}}, \bibinfo {author}
  {\bibfnamefont {J.~P.}\ \bibnamefont {Meyburg}}, \bibinfo {author}
  {\bibfnamefont {D.}~\bibnamefont {Diesing}}, \bibinfo {author} {\bibfnamefont
  {V.}~\bibnamefont {Roddatis}}, \bibinfo {author} {\bibfnamefont
  {I.}~\bibnamefont {Rungger}}, \bibinfo {author} {\bibfnamefont
  {M.}~\bibnamefont {Stamenova}}, \bibinfo {author} {\bibfnamefont
  {S.}~\bibnamefont {Sanvito}}, \bibinfo {author} {\bibfnamefont
  {U.}~\bibnamefont {Bovensiepen}}, \ and\ \bibinfo {author} {\bibfnamefont
  {A.}~\bibnamefont {Melnikov}},\ }\href@noop {} {\bibfield  {journal}
  {\bibinfo  {journal} {Physical Review Letters}\ }\textbf {\bibinfo {volume}
  {119}},\ \bibinfo {pages} {017202} (\bibinfo {year} {2017})}\BibitemShut
  {NoStop}%
\bibitem [{\citenamefont {Nenno}\ \emph {et~al.}(2016)\citenamefont {Nenno},
  \citenamefont {Kaltenborn},\ and\ \citenamefont {Schneider}}]{Nenno:2016cm}%
  \BibitemOpen
  \bibfield  {author} {\bibinfo {author} {\bibfnamefont {D.~M.}\ \bibnamefont
  {Nenno}}, \bibinfo {author} {\bibfnamefont {S.}~\bibnamefont {Kaltenborn}}, \
  and\ \bibinfo {author} {\bibfnamefont {H.~C.}\ \bibnamefont {Schneider}},\
  }\href@noop {} {\bibfield  {journal} {\bibinfo  {journal} {Physical Review
  B}\ }\textbf {\bibinfo {volume} {94}},\ \bibinfo {pages} {115102} (\bibinfo
  {year} {2016})}\BibitemShut {NoStop}%
\bibitem [{\citenamefont {Battiato}\ and\ \citenamefont
  {Held}(2016)}]{Battiato:2016bi}%
  \BibitemOpen
  \bibfield  {author} {\bibinfo {author} {\bibfnamefont {M.}~\bibnamefont
  {Battiato}}\ and\ \bibinfo {author} {\bibfnamefont {K.}~\bibnamefont
  {Held}},\ }\href@noop {} {\bibfield  {journal} {\bibinfo  {journal} {Physical
  Review Letters}\ }\textbf {\bibinfo {volume} {116}},\ \bibinfo {pages}
  {196601} (\bibinfo {year} {2016})}\BibitemShut {NoStop}%
\bibitem [{\citenamefont {Shokeen}\ \emph {et~al.}(2017)\citenamefont
  {Shokeen}, \citenamefont {Sanchez~Piaia}, \citenamefont {Bigot},
  \citenamefont {M{\"u}ller}, \citenamefont {Elliott}, \citenamefont
  {Dewhurst}, \citenamefont {Sharma},\ and\ \citenamefont
  {Gross}}]{Shokeen:2017uu}%
  \BibitemOpen
  \bibfield  {author} {\bibinfo {author} {\bibfnamefont {V.}~\bibnamefont
  {Shokeen}}, \bibinfo {author} {\bibfnamefont {M.}~\bibnamefont
  {Sanchez~Piaia}}, \bibinfo {author} {\bibfnamefont {J.~Y.}\ \bibnamefont
  {Bigot}}, \bibinfo {author} {\bibfnamefont {T.}~\bibnamefont {M{\"u}ller}},
  \bibinfo {author} {\bibfnamefont {P.}~\bibnamefont {Elliott}}, \bibinfo
  {author} {\bibfnamefont {J.~K.}\ \bibnamefont {Dewhurst}}, \bibinfo {author}
  {\bibfnamefont {S.}~\bibnamefont {Sharma}}, \ and\ \bibinfo {author}
  {\bibfnamefont {E.~K.~U.}\ \bibnamefont {Gross}},\ }\href@noop {} {\bibfield
  {journal} {\bibinfo  {journal} {Physical Review Letters}\ }\textbf {\bibinfo
  {volume} {119}},\ \bibinfo {pages} {1483} (\bibinfo {year}
  {2017})}\BibitemShut {NoStop}%
\bibitem [{\citenamefont {Zhu}\ \emph {et~al.}(2014)\citenamefont {Zhu},
  \citenamefont {Xu},\ and\ \citenamefont {Geng}}]{Zhu2014-oy}%
  \BibitemOpen
  \bibfield  {author} {\bibinfo {author} {\bibfnamefont {Y.-H.}\ \bibnamefont
  {Zhu}}, \bibinfo {author} {\bibfnamefont {D.-H.}\ \bibnamefont {Xu}}, \ and\
  \bibinfo {author} {\bibfnamefont {A.-C.}\ \bibnamefont {Geng}},\ }\href@noop
  {} {\bibfield  {journal} {\bibinfo  {journal} {Physica B: Physics of
  Condensed Matter}\ }\textbf {\bibinfo {volume} {446}},\ \bibinfo {pages} {43}
  (\bibinfo {year} {2014})}\BibitemShut {NoStop}%
\bibitem [{\citenamefont {{Faghihi}}\ \emph {et~al.}(2017)\citenamefont
  {{Faghihi}}, \citenamefont {{Carey}}, \citenamefont {{Michoski}},
  \citenamefont {{Hager}}, \citenamefont {{Janhunen}}, \citenamefont
  {{Chang}},\ and\ \citenamefont {{Moser}}}]{Faghihi:2017vv}%
  \BibitemOpen
  \bibfield  {author} {\bibinfo {author} {\bibfnamefont {D.}~\bibnamefont
  {{Faghihi}}}, \bibinfo {author} {\bibfnamefont {V.}~\bibnamefont {{Carey}}},
  \bibinfo {author} {\bibfnamefont {C.}~\bibnamefont {{Michoski}}}, \bibinfo
  {author} {\bibfnamefont {R.}~\bibnamefont {{Hager}}}, \bibinfo {author}
  {\bibfnamefont {S.}~\bibnamefont {{Janhunen}}}, \bibinfo {author}
  {\bibfnamefont {C.-S.}\ \bibnamefont {{Chang}}}, \ and\ \bibinfo {author}
  {\bibfnamefont {R.}~\bibnamefont {{Moser}}},\ }\href@noop {} {\bibfield
  {journal} {\bibinfo  {journal} {ArXiv e-prints}\ } (\bibinfo {year}
  {2017})},\ \Eprint {http://arxiv.org/abs/1702.05198} {arXiv:1702.05198
  [physics.plasm-ph]} \BibitemShut {NoStop}%
\bibitem [{\citenamefont {Strang}(1968)}]{Strang:1968dr}%
  \BibitemOpen
  \bibfield  {author} {\bibinfo {author} {\bibfnamefont {G.}~\bibnamefont
  {Strang}},\ }\href@noop {} {\bibfield  {journal} {\bibinfo  {journal} {SIAM
  Journal on Numerical Analysis}\ }\textbf {\bibinfo {volume} {5}},\ \bibinfo
  {pages} {506} (\bibinfo {year} {1968})}\BibitemShut {NoStop}%
\bibitem [{\citenamefont {Metzler}\ and\ \citenamefont
  {Klafter}(2000)}]{Metzler:2000hea}%
  \BibitemOpen
  \bibfield  {author} {\bibinfo {author} {\bibfnamefont {R.}~\bibnamefont
  {Metzler}}\ and\ \bibinfo {author} {\bibfnamefont {J.}~\bibnamefont
  {Klafter}},\ }\href@noop {} {\bibfield  {journal} {\bibinfo  {journal}
  {Physics reports}\ }\textbf {\bibinfo {volume} {339}},\ \bibinfo {pages} {1}
  (\bibinfo {year} {2000})}\BibitemShut {NoStop}%
\bibitem [{\citenamefont {Weiss}(2002)}]{Weiss:2002bb}%
  \BibitemOpen
  \bibfield  {author} {\bibinfo {author} {\bibfnamefont {G.~H.}\ \bibnamefont
  {Weiss}},\ }\href@noop {} {\bibfield  {journal} {\bibinfo  {journal} {Physica
  A: Statistical Mechanics and its Applications}\ }\textbf {\bibinfo {volume}
  {311}},\ \bibinfo {pages} {381} (\bibinfo {year} {2002})}\BibitemShut
  {NoStop}%
\bibitem [{\citenamefont {Battiato}\ \emph {et~al.}(2012)\citenamefont
  {Battiato}, \citenamefont {Carva},\ and\ \citenamefont
  {Oppeneer}}]{Battiato:2012hw}%
  \BibitemOpen
  \bibfield  {author} {\bibinfo {author} {\bibfnamefont {M.}~\bibnamefont
  {Battiato}}, \bibinfo {author} {\bibfnamefont {K.}~\bibnamefont {Carva}}, \
  and\ \bibinfo {author} {\bibfnamefont {P.~M.}\ \bibnamefont {Oppeneer}},\
  }\href@noop {} {\bibfield  {journal} {\bibinfo  {journal} {Physical Review
  B}\ }\textbf {\bibinfo {volume} {86}},\ \bibinfo {pages} {024404} (\bibinfo
  {year} {2012})}\BibitemShut {NoStop}%
\bibitem [{\citenamefont {Cao}\ \emph {et~al.}(1998)\citenamefont {Cao},
  \citenamefont {Gao}, \citenamefont {Elsayed-Ali}, \citenamefont {Miller},\
  and\ \citenamefont {Mantell}}]{Cao:1998ec}%
  \BibitemOpen
  \bibfield  {author} {\bibinfo {author} {\bibfnamefont {J.}~\bibnamefont
  {Cao}}, \bibinfo {author} {\bibfnamefont {Y.}~\bibnamefont {Gao}}, \bibinfo
  {author} {\bibfnamefont {H.~E.}\ \bibnamefont {Elsayed-Ali}}, \bibinfo
  {author} {\bibfnamefont {R.~J.~D.}\ \bibnamefont {Miller}}, \ and\ \bibinfo
  {author} {\bibfnamefont {D.~A.}\ \bibnamefont {Mantell}},\ }\href@noop {}
  {\bibfield  {journal} {\bibinfo  {journal} {Physical Review B}\ }\textbf
  {\bibinfo {volume} {58}},\ \bibinfo {pages} {10948} (\bibinfo {year}
  {1998})}\BibitemShut {NoStop}%
\bibitem [{\citenamefont {Melnikov}\ \emph
  {et~al.}(2011{\natexlab{b}})\citenamefont {Melnikov}, \citenamefont
  {Razdolski}, \citenamefont {Wehling}, \citenamefont {Papaioannou},
  \citenamefont {Roddatis}, \citenamefont {Fumagalli}, \citenamefont
  {Aktsipetrov}, \citenamefont {Lichtenstein},\ and\ \citenamefont
  {Bovensiepen}}]{Melnikov:2011epa}%
  \BibitemOpen
  \bibfield  {author} {\bibinfo {author} {\bibfnamefont {A.}~\bibnamefont
  {Melnikov}}, \bibinfo {author} {\bibfnamefont {I.}~\bibnamefont {Razdolski}},
  \bibinfo {author} {\bibfnamefont {T.~O.}\ \bibnamefont {Wehling}}, \bibinfo
  {author} {\bibfnamefont {E.~T.}\ \bibnamefont {Papaioannou}}, \bibinfo
  {author} {\bibfnamefont {V.}~\bibnamefont {Roddatis}}, \bibinfo {author}
  {\bibfnamefont {P.}~\bibnamefont {Fumagalli}}, \bibinfo {author}
  {\bibfnamefont {O.}~\bibnamefont {Aktsipetrov}}, \bibinfo {author}
  {\bibfnamefont {A.~I.}\ \bibnamefont {Lichtenstein}}, \ and\ \bibinfo
  {author} {\bibfnamefont {U.}~\bibnamefont {Bovensiepen}},\ }\href@noop {}
  {\bibfield  {journal} {\bibinfo  {journal} {Physical Review Letters}\
  }\textbf {\bibinfo {volume} {107}},\ \bibinfo {pages} {076601} (\bibinfo
  {year} {2011}{\natexlab{b}})}\BibitemShut {NoStop}%
\bibitem [{\citenamefont {Koopmans}\ \emph {et~al.}(2000)\citenamefont
  {Koopmans}, \citenamefont {van Kampen}, \citenamefont {Kohlhepp},\ and\
  \citenamefont {de~Jonge}}]{PhysRevLett.85.844}%
  \BibitemOpen
  \bibfield  {author} {\bibinfo {author} {\bibfnamefont {B.}~\bibnamefont
  {Koopmans}}, \bibinfo {author} {\bibfnamefont {M.}~\bibnamefont {van
  Kampen}}, \bibinfo {author} {\bibfnamefont {J.~T.}\ \bibnamefont {Kohlhepp}},
  \ and\ \bibinfo {author} {\bibfnamefont {W.~J.~M.}\ \bibnamefont
  {de~Jonge}},\ }\href {\doibase 10.1103/PhysRevLett.85.844} {\bibfield
  {journal} {\bibinfo  {journal} {Phys. Rev. Lett.}\ }\textbf {\bibinfo
  {volume} {85}},\ \bibinfo {pages} {844} (\bibinfo {year} {2000})}\BibitemShut
  {NoStop}%
\bibitem [{\citenamefont {Soulen}\ \emph {et~al.}(1998)\citenamefont {Soulen},
  \citenamefont {Byers}, \citenamefont {Osofsky}, \citenamefont {Nadgorny},
  \citenamefont {Ambrose}, \citenamefont {Cheng}, \citenamefont {Broussard},
  \citenamefont {Tanaka}, \citenamefont {Nowak}, \citenamefont {Moodera},
  \citenamefont {Barry},\ and\ \citenamefont {Coey}}]{Soulen85}%
  \BibitemOpen
  \bibfield  {author} {\bibinfo {author} {\bibfnamefont {R.~J.}\ \bibnamefont
  {Soulen}}, \bibinfo {author} {\bibfnamefont {J.~M.}\ \bibnamefont {Byers}},
  \bibinfo {author} {\bibfnamefont {M.~S.}\ \bibnamefont {Osofsky}}, \bibinfo
  {author} {\bibfnamefont {B.}~\bibnamefont {Nadgorny}}, \bibinfo {author}
  {\bibfnamefont {T.}~\bibnamefont {Ambrose}}, \bibinfo {author} {\bibfnamefont
  {S.~F.}\ \bibnamefont {Cheng}}, \bibinfo {author} {\bibfnamefont {P.~R.}\
  \bibnamefont {Broussard}}, \bibinfo {author} {\bibfnamefont {C.~T.}\
  \bibnamefont {Tanaka}}, \bibinfo {author} {\bibfnamefont {J.}~\bibnamefont
  {Nowak}}, \bibinfo {author} {\bibfnamefont {J.~S.}\ \bibnamefont {Moodera}},
  \bibinfo {author} {\bibfnamefont {A.}~\bibnamefont {Barry}}, \ and\ \bibinfo
  {author} {\bibfnamefont {J.~M.~D.}\ \bibnamefont {Coey}},\ }\href {\doibase
  10.1126/science.282.5386.85} {\bibfield  {journal} {\bibinfo  {journal}
  {Science}\ }\textbf {\bibinfo {volume} {282}},\ \bibinfo {pages} {85}
  (\bibinfo {year} {1998})}\BibitemShut {NoStop}%
\bibitem [{Note1()}]{Note1}%
  \BibitemOpen
  \bibinfo {note} {We have checked this using standard optical-transfer-matrix
  calculations.}\BibitemShut {Stop}%
\bibitem [{\citenamefont {Hofherr}\ \emph
  {et~al.}(2017{\natexlab{b}})\citenamefont {Hofherr}, \citenamefont
  {Maldonado}, \citenamefont {Schmitt}, \citenamefont {Berritta}, \citenamefont
  {Bierbrauer}, \citenamefont {Sadashivaiah}, \citenamefont {Schellekens},
  \citenamefont {Koopmans}, \citenamefont {Steil}, \citenamefont {Cinchetti},
  \citenamefont {Stadtm{\"u}ller}, \citenamefont {Oppeneer}, \citenamefont
  {Mathias},\ and\ \citenamefont {Aeschlimann}}]{Hofherr:2017dy}%
  \BibitemOpen
  \bibfield  {author} {\bibinfo {author} {\bibfnamefont {M.}~\bibnamefont
  {Hofherr}}, \bibinfo {author} {\bibfnamefont {P.}~\bibnamefont {Maldonado}},
  \bibinfo {author} {\bibfnamefont {O.}~\bibnamefont {Schmitt}}, \bibinfo
  {author} {\bibfnamefont {M.}~\bibnamefont {Berritta}}, \bibinfo {author}
  {\bibfnamefont {U.}~\bibnamefont {Bierbrauer}}, \bibinfo {author}
  {\bibfnamefont {S.}~\bibnamefont {Sadashivaiah}}, \bibinfo {author}
  {\bibfnamefont {A.~J.}\ \bibnamefont {Schellekens}}, \bibinfo {author}
  {\bibfnamefont {B.}~\bibnamefont {Koopmans}}, \bibinfo {author}
  {\bibfnamefont {D.}~\bibnamefont {Steil}}, \bibinfo {author} {\bibfnamefont
  {M.}~\bibnamefont {Cinchetti}}, \bibinfo {author} {\bibfnamefont
  {B.}~\bibnamefont {Stadtm{\"u}ller}}, \bibinfo {author} {\bibfnamefont
  {P.~M.}\ \bibnamefont {Oppeneer}}, \bibinfo {author} {\bibfnamefont
  {S.}~\bibnamefont {Mathias}}, \ and\ \bibinfo {author} {\bibfnamefont
  {M.}~\bibnamefont {Aeschlimann}},\ }\href@noop {} {\bibfield  {journal}
  {\bibinfo  {journal} {Physical Review B}\ }\textbf {\bibinfo {volume} {96}},\
  \bibinfo {pages} {100403} (\bibinfo {year} {2017}{\natexlab{b}})}\BibitemShut
  {NoStop}%
\end{thebibliography}

\end{document}